\documentclass{article}
\usepackage{graphicx} 
\usepackage{subcaption}
\usepackage{authblk}
\usepackage{bm}
\usepackage{mathrsfs}
\usepackage{amsbsy}
\usepackage{amsmath}
\usepackage{amssymb}
\usepackage{braket}
\usepackage[numbers,sort&compress]{natbib}
\usepackage{lipsum}
\usepackage[]{hyperref}
\usepackage{booktabs} 
\usepackage{multirow}
\usepackage[english]{babel}
\usepackage[letterpaper, textwidth=6in, textheight=9in]{geometry}
\usepackage[labelformat=simple]{subcaption}

\DeclareUnicodeCharacter{2212}{-}
\usepackage{indentfirst}
\begin{document}

\begin{center}

{\bf \Large{Holographic fermions in the Dyonic Gubser-Rocha black hole}}\\	
\vspace{1.6cm}	
{\small{\textbf{Cheng-Yuan Lu}$^{1}$, ~\textbf{Xian-Hui Ge}$^{1,2*}$},~\textbf{Sang-Jin Sin}$^{3}$}\let\thefootnote\relax
\footnotetext{* Corresponding author. gexh@shu.edu.cn}\\

		\vspace{0.8cm}
		
		$^1${\it Department of Physics, Shanghai University, Shanghai 200444,  China} \\
            $^2${\it Shanghai Key Laboratory of High Temperature Superconductors, Shanghai 200444,  China} \\
            $^3${\it Department of Physics, Hanyang University, Seoul, 04763, South Korea}

		\vspace{1.6cm}

    \begin{abstract}
	We investigate the fermionic properties of a dyonic Gubser-Rocha model in the context of gauge/gravity duality. This model incorporates both a magnetic field and momentum relaxation. We have derived this model's scaling exponent, revealing the influence of the magnetic field and momentum relaxation on low-energy physics. As the magnetic field strength and momentum relaxation increase, the spectral function of the dual field changes significantly. Specifically, we observe variations in the scaling exponent, Fermi momentum, and dispersion relations as the magnetic field increases, highlighting the system's transition from a Fermi liquid to a non-Fermi liquid, and eventually to an insulating state. Our analysis of the magneto-scattering rate reveals that it is nearly zero in the Fermi liquid region, increases significantly in the non-Fermi liquid region, and ultimately arrives at a maximum value in the insulating state.
    \end{abstract}
  
\end{center}
\newpage
\tableofcontents
\newpage

\section{Introduction}

The Landau Fermi liquid theory has been highly successful in describing quantum many-body fermionic systems, particularly at low temperatures, where strongly interacting Fermi liquids exhibit behavior that can be captured by quasi-particles with effective masses and spin interactions \cite{Landau:1958joj}. These quasi-particles obey Fermi-Dirac statistics and form the foundation of Fermi liquid theory. However, in the 1980s, the discovery of new materials, particularly high-temperature cuprate superconductors \cite{bednorz1986possible}, revealed properties that extend beyond the quasi-particle framework. These materials display unusual thermodynamic and transport behaviors, such as T-linear resistance, which cannot be adequately explained by Fermi liquid theory, thus giving rise to what is now referred to as non-Fermi liquids.

In 1997, Juan Maldacena introduced the groundbreaking Anti-de Sitter/Conformal Field Theory (AdS/CFT) Correspondence, by analyzing the $\mathcal{N}=4$ Yang-Mills theory in an $AdS_{5} \times S^5$ spacetime \cite{Maldacena:1997re}. This duality opened a novel approach to studying strongly coupled systems. While gravity remains weakly coupled, its dual counterpart can capture the dynamics of strongly coupled systems. In 2009, Liu et al. demonstrated the possibility of exploring fermionic systems by introducing a probe Fermi field into the dual bulk spacetime \cite{liu2011non}. They extracted the Fermi response of a condensed matter system from a Reissner-Nordström (RN)-AdS black hole, corresponding to a non-Fermi liquid. Through this approach, the spectral function, a key observable in angle-resolved photoemission spectroscopy (ARPES) experiments, could be obtained.

The gauge/gravity duality has since provided a powerful framework for studying strongly coupled fermionic systems, extending beyond the quasi-particle picture and encompassing Fermi liquids, non-Fermi liquids, and marginal Fermi liquids. Early studies focused on the Reissner-Nordström AdS background \cite{liu2011non,faulkner2011emergent,_ubrovi__2009}, revealing corrections to transport properties due to holographic fermions \cite{doi:10.1126/science.1189134}. These studies also showed how holographic fermions could lead to Cooper pairing and superconducting instabilities \cite{Hartman_2010}. Subsequent work expanded this framework to include magnetic field backgrounds \cite{Gubankova:2010ny,gubankova2011holographic,Ge:2011cw}, momentum relaxation \cite{Fang:2015dia,Jeong:2019zab}, holographic superconductors \cite{Yuk:2022lof,DeWolfe:2016rxk,ghorai2024order,ghorai2023fermi}, anisotropic backgrounds \cite{Fang:2014jka,chakrabarti2022studyingholographicfermisurface}, holographic lattices \cite{Ling:2014bda,Ling:2013aya,han2024mean,yuk2024encoding}, dipole couplings \cite{li2012holographic,wen2012dipole}, the Weyl semi-metal \cite{song2019stability}, and the Mott transition \cite{Seo:2018hrc,ghorai2024classes}. Despite its successes, the holographic Green's function does not satisfy the electronic sum rule, prompting the development of semi-holography as a more effective theoretical framework \cite{Gursoy:2011gz, Faulkner:2010tq}. Recently, holographic fermions have been applied in experimental contexts \cite{Smit:2021dwh, Mauri:2024wgc}.

The Gubser-Rocha model, proposed in 2009 \cite{Gubser:2009qt}, offers a compelling description of linear resistivity in strange metals. Strange metals are classified as non-Fermi liquids and known to be the normal phase of high-temperature superconductors. Investigations into the Fermi response of the Gubser-Rocha model \cite{li2012holographic,wu2011some,gubser2012analytic,wen2012dipole} have shown that while the dual boundary fermionic system shares similarities with Fermi liquids, it does not fully adhere to the conventional properties of Fermi liquids.

More recently, a new Gubser-Rocha model incorporating S-duality was introduced \cite{ge2024thermo}. This extended version of the (3+1)-dimensional Gubser-Rocha model \cite{Gubser:2009qt} includes both a magnetic field and momentum relaxation terms. The thermoelectric transport properties of this model, which are closely tied to the behavior of electrons near the Fermi surface, have been studied \cite{ge2024thermo, Ishigaki:2024pfv}. In this work, we aim to explore the Fermi response of this extended system, with a particular focus on how the magnetic field influences its behavior. Additionally, we investigate how low-energy physics describes the magnetic effects.

This paper is organized as follows: In Section 2, we review the dyonic Gubser-Rocha model. In Section 3, we analyze the IR physics, deriving the scaling exponent and the IR Green's function. Section 4 presents the spectral function for various Landau levels, considering different magnetic field strengths and momentum relaxation parameters. In Section 5, we compute the Fermi momentum, scaling exponent, and dispersion relation by numerically solving the Dirac equation. Section 6 discusses the scattering rate, drawing on both UV and IR results. Finally, Section 8 summarizes the main contributions of this paper.

\section{Holographic Fermions}
\subsection{Set up}
We consider the Dyonic Gubser-Rocha model given in \cite{ge2024thermo}. Its action is given by
\begin{align}
    S &= \frac{1}{2 \kappa^2} \int \mathrm{d}^4 x\left(\mathcal{L}_1+\mathcal{L}_2\right),\\
    \frac{\mathcal{L}_1}{\sqrt{-g}} &= R-\frac{3}{2} \frac{\partial_\mu \tau \partial^\mu \bar{\tau}}{(\operatorname{Im} \tau)^2}-\frac{1}{4} e^{-\phi} F^2+\frac{1}{4} \chi F \tilde{F}+\frac{3}{L^2} \frac{\tau \bar{\tau}+1}{\operatorname{Im} \tau},\\
    \frac{\mathcal{L}_2}{\sqrt{-g}} &= -\frac{1}{2} \sum_{I=1,2}\left(\partial \psi_I\right)^2,
\end{align}
where $\mathcal{L}_1$ is the Lagrangian density of the S-completed Gubser-Rocha model, $\mathcal{L}_2$ is the Lagrangian density of a linear axion model, and $\tau$ is the axio-dilaton. We set the gravitational constant $\kappa$ and the AdS radius $L$ to $2\kappa^2 = L = 1$.

The model features the following gravitational solution, where the metric is expressed as
\begin{align}
 \mathrm{d} s^2 &= -f(r) \mathrm{d} t^2+\frac{\mathrm{d} r^2}{f(r)}+h(r)\left(\mathrm{d} x^2+\mathrm{d} y^2\right), \\
 f(r) &= h(r)\left(1-\frac{n^2+B^2}{3 \rho\left(\rho+r\right)^3}-\frac{ \beta^2}{2 \left(\rho+r\right)^2}\right), \\
 h(r) &= \sqrt{r\left(r+\rho\right)^3},
 \label{fr}\\
 A &= n\left(\frac{1}{r_0+\rho}-\frac{1}{r+\rho}\right) \mathrm{d} t-B y \mathrm{~d} x,\label{A}\\
 \psi_I&=\beta x_I,
\end{align}
where $n$ represents the charge density, $B$ represents an external magnetic field and $\beta$ denotes the momentum relaxation term. $\rho$ is a real parameter determining the location of the curvature
singularity at $r=-\rho$. $r_0$ is the outermost horizon radius defined by $f(r_0) = 0$.

From \cite{Ishigaki:2024pfv}, we know the chemical potential of the boundary field and the temperature is given by
\begin{align}
\mu=\frac{n}{r_0+\rho},\,\,T=\frac{r_0}{8 \pi \sqrt{r_0 ( r_0 + \rho )^3}} (6(r_0+\rho)^2-\beta^2).
\label{uT}
\end{align}
By solving the equations, we can express $r_0$ and $\rho$ as follows,
\begin{align}
  \frac{r_0}{\mu}=\frac{64 \pi^2 n^3  T^2  }{ ( 6 n^2- \mu^2 \beta^2)^2},\,\,\frac{\rho}{\mu}=\frac{n(36 n^4-12   \mu^2 \beta^2 n^2 -64  \pi^2\mu^2 n^2 T^2 + \mu^4 \beta^4)}{\mu^2 ( \mu^2 \beta^2-6 n^2)^2} .
  \label{r0}
\end{align}

In this study, we fix the chemical potential and perform the rescaling, 
$n=n \mu^2, T=T \mu,\beta=\beta \mu,r=r \mu.$ By substituting the expressions for $r_0$ and $\rho$ into the metric, we obtain the metric for the black hole in the rescaled variables as 
\begin{align}
 g_{tt}=&-\sqrt{r(r+n-\frac{64 n^3 \pi^2 T^2}{(\beta^2-6n^2)^2})^3}\nonumber\\
 &\left(1 + \frac{n(\beta^2-6n^2)^6(\beta^2-2n^2)}{2\left((r+n)(\beta^2-6n^2)^2-64\pi^2n^3T^2\right)^3} - \frac{\beta^{2}}{2\left(r+n-\frac{64\pi^2n^3T^2}{(-6n^2+\beta^2)^2}\right)^2}\right) ,\label{gtt}\\ g_{rr}=&-{g_{tt}}^{-1},\\g_{xx}=&g_{yy}=\sqrt{r(r+n-\frac{64 n^3 \pi^2 T^2}{(\beta^2-6n^2)^2})^3} .
\end{align}
In this situation, the magnetic term $B$ is given by
\begin{align}
B=n\sqrt{\frac{1}{2}\left(-2-3\beta^2+6n^2\right)+\frac{96n^2(\beta^2-2n^2)\pi^2T^2}{(\beta^2-6n^2)^2}}.
\label{magnetic}
\end{align}

Next, we introduce a probe fermion in the gravitational background, considering the bulk fermion action 
\begin{align}
    S_{\rm{fermion}}=i \int d^{4} x \sqrt{-g} \bar{\zeta}\left(\Gamma^\alpha \mathcal{D}_\alpha-m\right) \zeta,
\end{align}
where $\mathcal{D}_\alpha=\partial_\alpha+\frac{1}{4}\left(\omega_{\mu \nu}\right)_\alpha \Gamma^{\mu \nu}-i q A_\alpha$  is the covariant derivative, incorporating the spin connection $\left(\omega_{\mu \nu}\right)_\alpha$. This leads to the Dirac equation
\begin{align}
    \Gamma^\alpha \mathcal{D}_\alpha \zeta-m \zeta=0.
    \label{Dirac}
\end{align}
The magnetic field dependence in the equation comes from the covariant derivative term in the interaction. This term introduces the coupling between the fermions and the electromagnetic field, and the magnetic field explicitly enters through the vector potential $A_\alpha$. From \eqref{A}, we have $A_t(r)=\mu(1-\frac{r_0+\rho}{r+\rho})$ and $A_x(y)=By$. Here, we apply Einstein's summation convention for the index $\alpha$. We can separate the $y$- and $r$-dependent components in the Dirac equation. For the $y$-dependent part, this corresponds exactly to the Schrödinger equation of the harmonic oscillator, yielding an exact solution \cite{gubankova2011holographic}. In the AdS/CFT duality, the
$r$-direction corresponds to the energy scale of the boundary field theory. We focus on the $r$-direction flow equation in the presence of a magnetic field, which is given by  
\begin{align}  
\sqrt{\frac{g_{i i}}{g_{r r}}} \partial_r \xi_{ \pm}^{(l)}=-2 m \sqrt{g_{i i}} \xi_{ \pm}^{(l)}+\left(u(r) \pm \lambda_l\right)^2\left(\xi_{ \pm}^{(l)}\right)^2+\left(u(r) \mp \lambda_l\right).
\label{flow}
\end{align}
Here, $\lambda_l=\sqrt{2|q B| l}$ represents the discrete effective momentum, and $l$ denotes the Landau level \cite{gubankova2011holographic}. Applying the boundary condition $\xi |_{r=0}=i$, we can derive Green's function in the boundary field
\begin{align}
G_R(\omega, l)&=\left.\lim _{\epsilon \rightarrow 0} \epsilon^{-2 m}\left(\begin{array}{cc}
\xi_{+}^{(l)} & 0 \\
0 & \xi_{-}^{(l)}
\end{array}\right)\right|_{r=\frac{1}{\epsilon}},
\label{G^-1}
\end{align}
and the spectral function
\begin{align}
A\left(\omega,k_x, l_y\right)&=\text{Tr}[\text{Im } G_R(\omega,k_x, l_y)].
\end{align}
We can extract the fermion response of the boundary field by solving the bulk Dirac equation \eqref{Dirac}. The spectral function serves as an observable in angle-resolved photoemission spectroscopy. It can be a bridge between the holographic principle and the experiments in the real world. 

\subsection{Dirac equation and infrared analysis}
The holographic principle suggests that the near-horizon geometry of the bulk governs the infrared (IR) physics of the boundary field. The imaginary part of the self-energy is entirely determined by the imaginary part of
the infrared Green's function at low temperatures and energies \cite{Mauri:2024wgc}. Thus, we will conduct an IR analysis of our model.

Returning to the Dirac equation, we can eliminate the covariant derivative in \eqref{Dirac} after rescaling $\zeta=\left(-g g^{r r}\right)^{-\frac{1}{4}}\Psi$, 
\begin{align}    \left[\Gamma^\alpha\left(\partial_\alpha-i q A_\alpha\right)-m\right] \Psi=0,
\end{align}  
where $\Psi$ is the four-component spinor. We choose the Gamma matrices as in \cite{gubser2012analytic} and perform a Fourier transformation. The Dirac equation simplifies to
\begin{align}
 [\sqrt{-g^{t t}} \sigma_1\left(\omega+q A_t\right)+\sqrt{g^{r r}} \sigma_3 \partial_r+(-1)^\alpha \sqrt{g^{x x}} i \sigma_2 k-m] \psi_\alpha=0.
 \label{pauli}
\end{align}
The Dirac equation is decoupled into two equations where $\alpha=1,2.$  $\psi_1$ and $\psi_2$ represent the two-component spinors. By analyzing \eqref{pauli}, we can readily observe that $\psi_1 (k,\omega)=\psi_2 (-k,\omega)$. So focusing on one of them is enough. By substituting the representations of the Pauli matrices into the equation \eqref{pauli}, the Dirac equation takes the form
\begin{align}
\partial_r \psi_2 =\left(\begin{array}{cc}
m \sqrt{g_{r r}} & \sqrt{\frac{g_{r r}}{g_{x x}}} k-\sqrt{\frac{g_{r r}}{-g_{t t}}}\left(\omega+q A_t\right) \\
\sqrt{\frac{g_{r r}}{g_{x x}}} k+\sqrt{\frac{g_{r r}}{-g_{t t}}}\left(\omega+q A_t\right) & -m \sqrt{g_{r r}}
\end{array}\right)\psi_2.
\end{align}

Next, we investigate the low-temperature near-horizon metric. Expanding $g_{tt}$ in \eqref{gtt} in powers of $r$ at low temperatures, we obtain 
\begin{align}
g_{tt} \approx -\frac{(-\beta^2+6n^2)r^{3/2}}{2\sqrt{n^3}L^2}\left(1-\frac{64n^3\pi^2T^2}{(-\beta^2+6n^2)^2r}\right).
\end{align}
Following a similar approach as in \cite{faulkner2011emergent}, we perform the coordinate transformation
\begin{align}
 r = \omega^2 \frac{4 L_2^4}{\xi^2},\,\,t=\frac{\tau}{\omega}  ,
\end{align}
This transformation leads to the conformal $\rm{AdS_2}$ black hole metric in the $\rm{IR}$ region,
\begin{align}
ds^2&=8\frac{\omega L^2}{a \xi}\left(\frac{L_2^2}{\xi^2} \left(-\left( 1 - \frac{4 \pi^2 T^2}{L^4} \frac{\xi^2}{\omega^2} \right) d\tau^2 + \frac{1}{\left( 1 - \frac{4 \pi^2 T^2}{L^4} \frac{\xi^2}{\omega^2} \right)} d\xi^2\right)+\frac{\sqrt{n^3}}{4L^2}(dx^2+dy^2)\right),\label{AdS_2}
\end{align}
where $a$ is a parameter that we have redefined given by 
\begin{align}
a&=\frac{6n^2-\beta^2}{2\sqrt{n^3}}.    
\end{align}

The IR geometry of the dyonic Gubser-Rocha model conforms to $\rm{AdS}_2$ black hole with $L_2=\frac{L}{\sqrt{a}}$, paralleling the Gubser-Rocha model discussed in \cite{gubser2012analytic}. 
From the above metric \eqref{AdS_2}, we can derive 
\begin{align}
g_{xx}=\frac{2\omega\sqrt{n^3}}{a \xi},g_{tt}=-a \, r^{3/2} \left(1 - \frac{16 \pi^2 T^2}{a^2 r}\right),g_{rr}=\frac{1}{a \, r^{3/2} \left(1 - \frac{16 \pi^2 T^2}{a^2 r}\right)} .
\label{g}
\end{align}
Substituting these expressions into the Dirac equation yields:
\begin{align}
\mathcal{M} \left(\begin{array}{cc} y_1(\xi) \\ y_2(\xi) \end{array}\right)=- \xi \sqrt{1 - \frac{4 \pi^2 T^2 \xi^2}{\omega^2}} \left(\begin{array}{cc} y_1^{'}(\xi) \\ y_2^{'}(\xi)\end{array}\right),\label{IR}
\end{align}
\begin{align}
 \mathcal{M} =\left(\begin{array}{cc}
\frac{2\sqrt{2} m}{a} \sqrt{\frac{\omega}{\xi}} & \frac{2 k}{\sqrt{a\sqrt{n^3} }} - \frac{1}{\sqrt{1-\frac{4 \pi^2 T^2 \xi^2}{\omega^2}}} \frac{\xi}{\omega} \left( \omega + q \frac{4 \omega^2}{n a^2 \xi^2} \right) \\
\frac{2k}{\sqrt{a\sqrt{n^3}}} + \frac{1}{\sqrt{1 - \frac{4\pi^2 T^2 \xi^2}{\omega^2}}} \frac{\xi}{\omega} \left( \omega + q \frac{4\omega^2}{na^2\xi^2} \right) & -\frac{2\sqrt{2} m}{a} \sqrt{\frac{\omega}{\xi}}
\end{array}\right).
\end{align}
Here, $\psi_2=\left(\begin{array}{cc} y_1 \\ y_2\end{array}\right) $. Notably, the dyonic Gubser-Rocha model is similar to the Gubser-Rocha model but different from the RN-AdS model. First, we focus on the zero-temperature case, $T=0$. The dependence of the coupled fermion's mass and charge term on $\frac{\omega}{\xi}$ is of small order. This follows the matching rule from \cite{faulkner2011emergent} $\xi \rightarrow 0$ and $\frac{\omega}{\xi}\rightarrow0$. In this limit, equation \eqref{IR} has two linearly independent solutions,
\begin{align}
y_1\left(\xi\right)&= \xi^{2 k \sqrt{\frac{1}{a \sqrt{n^3}}}} {C}_1+\xi^{-2 k\sqrt{\frac{1}{a \sqrt{n^3}}}} {C}_2, \label{solution1}\\
y_2\left(\xi\right) &=-\xi^{2 k \sqrt{\frac{1}{a \sqrt{n^3}}}} {C}_1+\xi^{-2 k\sqrt{\frac{1}{a \sqrt{n^3}}}} {C}_2 .  
\label{solution2}
\end{align}
The scaling exponent can be identified as
\begin{align}
\nu=  2 k \sqrt{\frac{2}{6n^2-\beta^2}}.  
\label{nu}
\end{align}
Here, $k$ denotes momentum, and $\beta$ represents momentum relaxation. When $k=k_F$, $\nu_{k_F}$ serves as a critical parameter for evaluating the properties of the boundary field in relation to the non-Fermi liquid. Specifically, the system exhibits Fermi liquid behavior when $\nu_{k_F}>\frac{1}{2}$, transitions to a marginal Fermi liquid at $\nu_{k_F}=\frac{1}{2}$, and displays non-Fermi liquid behavior when $\nu_{k_F}<\frac{1}{2}$. The scaling exponent in the dyonic RN-AdS model is more intricate, as it depends on the mass and charge parameters of the coupled fermion. In contrast, the scaling behavior in our model is determined solely by the magnetic field and momentum relaxation parameters. 

To generalize the solutions for $\xi \rightarrow 0$ and $\xi \rightarrow \infty$, we rewrite the equation \eqref{IR} as 
\begin{align}
\left(\begin{array}{cc}
0 & \nu-\xi \\
\nu+\xi & 0
\end{array}\right) \left(\begin{array}{cc} y_1(\xi) \\ y_2(\xi) \end{array}\right)=-\xi \left(\begin{array}{cc} y_1^{'}(\xi) \\ y_2^{'}(\xi)\end{array}\right) .
\end{align}
It is easy for us to handle this equation with the change of variables \cite{Mauri:2024wgc}
\begin{align}
\binom{y_1}{y_2}=\frac{1}{\sqrt{2}}\left(\begin{array}{cc}
1 & 1 \\
-i & i
\end{array}\right)\binom{u_1}{u_2}  .  
\end{align}
Consequently, we derive
\begin{align}
\begin{aligned}
& \partial_\xi^2 u_{1}+\frac{\partial_\xi u_{1}}{\xi}+u_{1}\left(\frac{i}{\xi}-\frac{\nu^2}{\xi^2}+1\right)=0 ,\\
& u_{2}=-\frac{\xi}{i \nu}\left(\partial_\xi u_{1}+i u_{1}\right).
\end{aligned}
\end{align}
Adopting the infalling boundary condition, we find that the solution approaches the asymptotic form $e^{i\xi}$ as $\xi \rightarrow \infty$. The solutions can be expressed as
\begin{align}
\begin{aligned}
& y_{1}(\xi)=\frac{C}{2 \sqrt{\xi}}\left(i \nu W_{-1 / 2, \nu}(-2 i \xi)+W_{1 / 2, \nu}(-2 i \xi)\right), \\
& y_{2}(\xi)=\frac{C}{2 \sqrt{\xi}}\left(\nu W_{-1 / 2, \nu}(-2 i \xi)+i W_{1 / 2, \nu}(-2 i \xi)\right).
\end{aligned}
\end{align}
Here, $W$ denotes the Whittaker function, and we revert back to the $y$ variables. Ultimately, we obtain the IR Green function by expanding the solutions as $\xi \rightarrow 0$,
\begin{align}
\mathcal{G}_{IR} =\frac{i (-4)^{-\nu} \Gamma\left[\frac{1}{2}-\nu\right]}{\Gamma\left[\frac{1}{2}+\nu\right]}(2 \omega)^{2 \nu}.   
\label{IR Green}
\end{align}
Similarly, in the non-zero temperature case, with $T \ll 1$, we generalize the IR Green function \eqref{IR Green} and find $\mathcal{G}_{IR}(T) \propto T^{2 \nu}$. The scattering rate of the quasiparticle corresponds to the imaginary part of the IR Green function given by
\begin{align}
\gamma \propto  \rm{Im} \left[ \mathcal{G}_{IR} \right] .
\label{gamma}
\end{align}

The form of the IR Green function aligns with that of the Gubser-Rocha model, albeit with a different scaling exponent. Notably, as both momentum relaxation and the magnetic field approaches zero, the scaling exponent \eqref{nu} would return to that of the original Gubser-Rocha model. This indicates that momentum relaxation and magnetic field effects significantly influence the IR physics in the dyonic Gubser-Rocha model.

\begin{figure}[htbp]
    \centering
    \begin{subfigure}{0.25\textwidth}
        \centering
        \includegraphics[width=\textwidth]{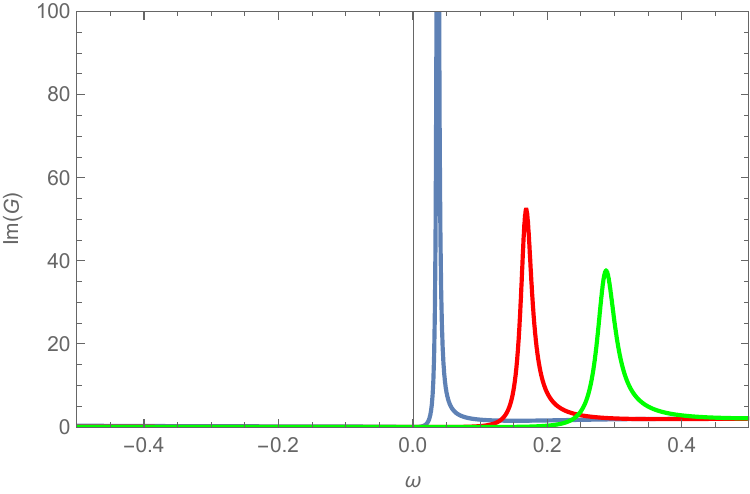}
        \caption{m=0  $\frac{B}{n}=0.2$}
        \label{}
    \end{subfigure}
    \begin{subfigure}{0.25\textwidth}
        \centering
        \includegraphics[width=\textwidth]{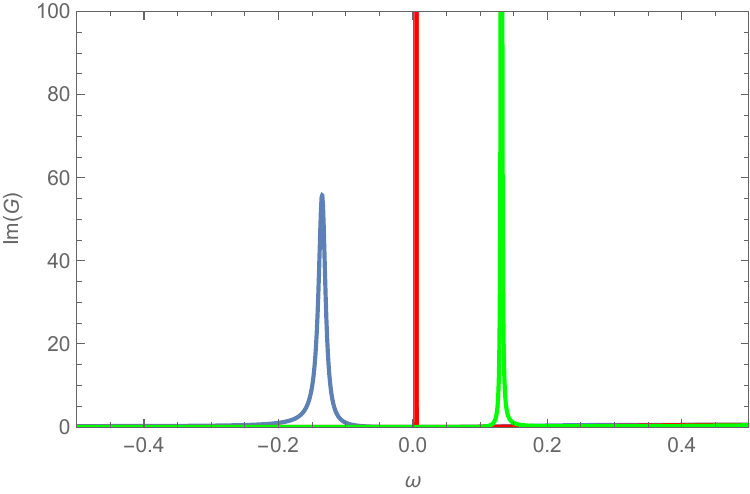}
        \caption{m=-0.25  $\frac{B}{n}=0.2$}
        \label{}
    \end{subfigure}
    \begin{subfigure}{0.25\textwidth}
        \centering
        \includegraphics[width=\textwidth]{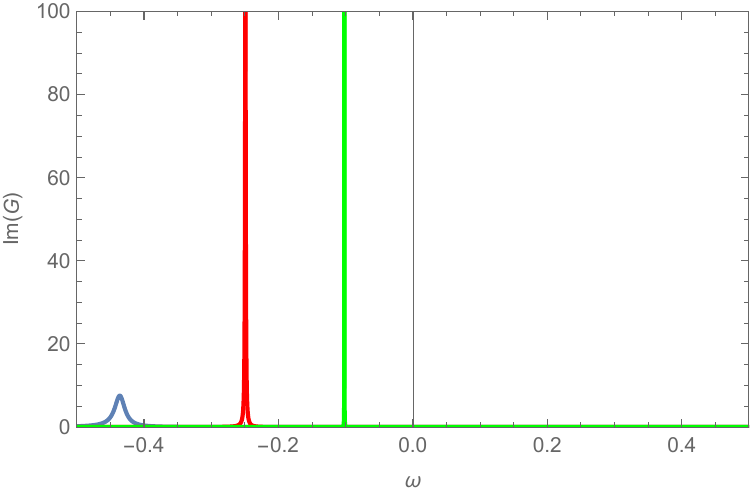}
        \caption{m=-0.49  $\frac{B}{n}=0.2$}
        \label{}
    \end{subfigure}
    
    \begin{subfigure}{0.25\textwidth}
        \centering
        \includegraphics[width=\textwidth]{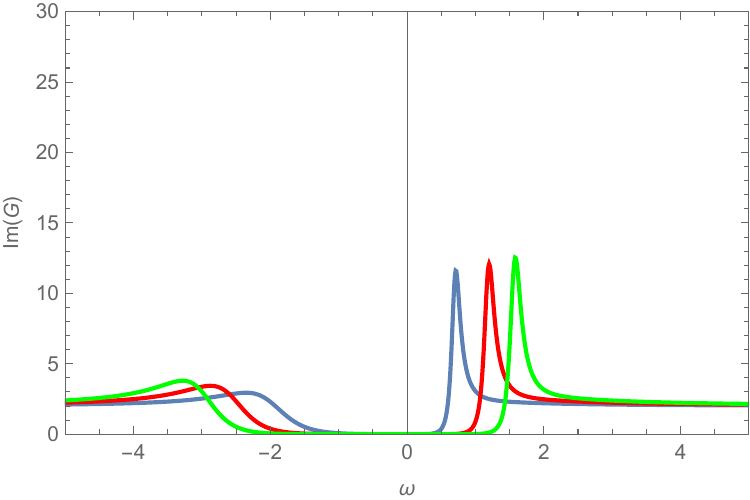}
        \caption{m=0 $\frac{B}{n}=1$}
        \label{}
    \end{subfigure}
    \begin{subfigure}{0.25\textwidth}
        \centering
        \includegraphics[width=\textwidth]{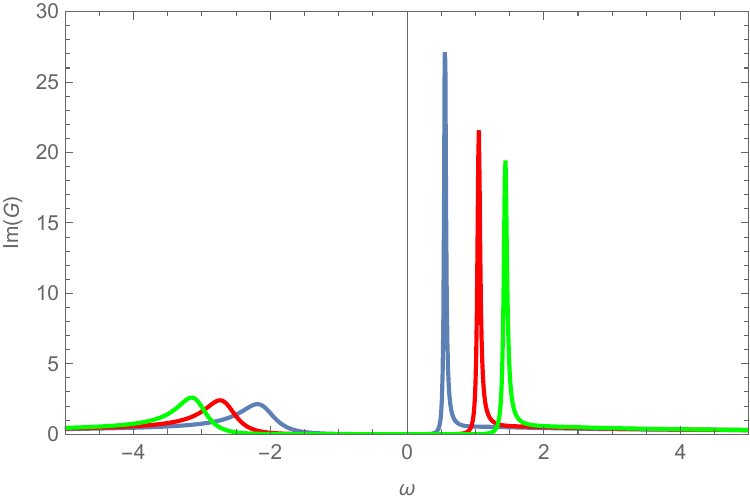}
        \caption{m=-0.25  $\frac{B}{n}=1$}
        \label{}
    \end{subfigure}
    \begin{subfigure}{0.25\textwidth}
        \centering
        \includegraphics[width=\textwidth]{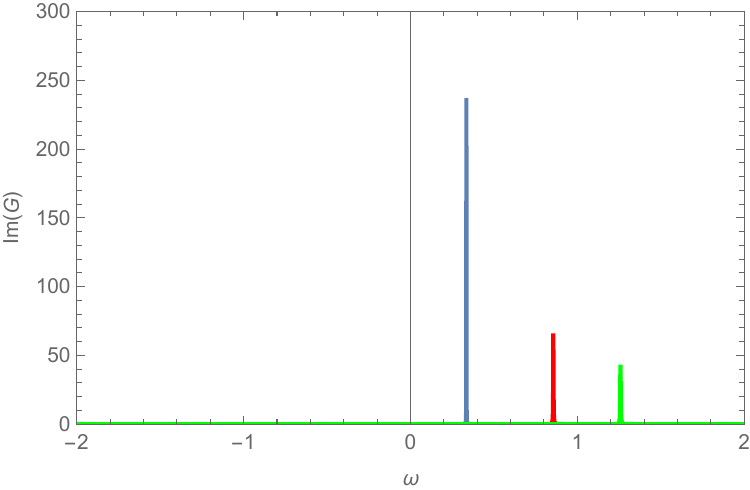}
        \caption{m=-0.49  $\frac{B}{n}=1$}
        \label{}
    \end{subfigure}
    
    \begin{subfigure}{0.25\textwidth}
        \centering
        \includegraphics[width=\textwidth]{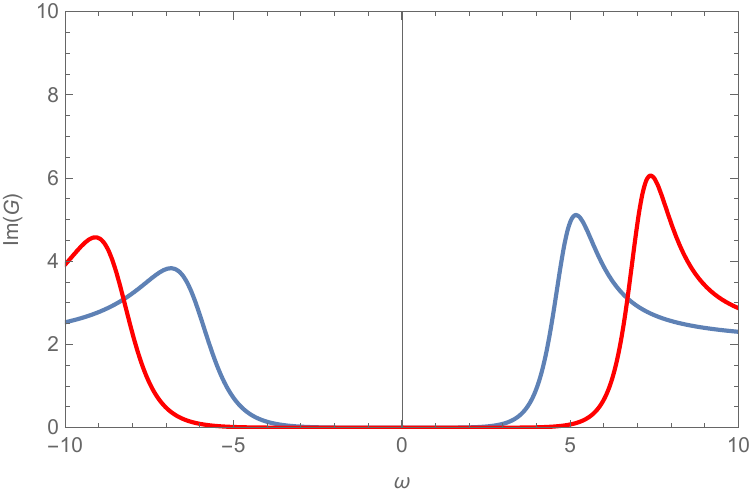}
        \caption{m=0  $\frac{B}{n}=5$}
        \label{}
    \end{subfigure}
    \begin{subfigure}{0.25\textwidth}
        \centering
        \includegraphics[width=\textwidth]{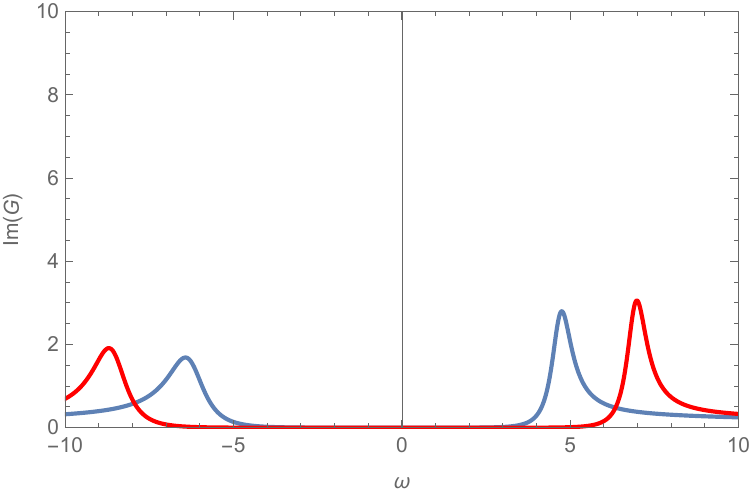}
        \caption{m=-0.25  $\frac{B}{n}=5$}
        \label{}
    \end{subfigure}
    \begin{subfigure}{0.25\textwidth}
        \centering
        \includegraphics[width=\textwidth]{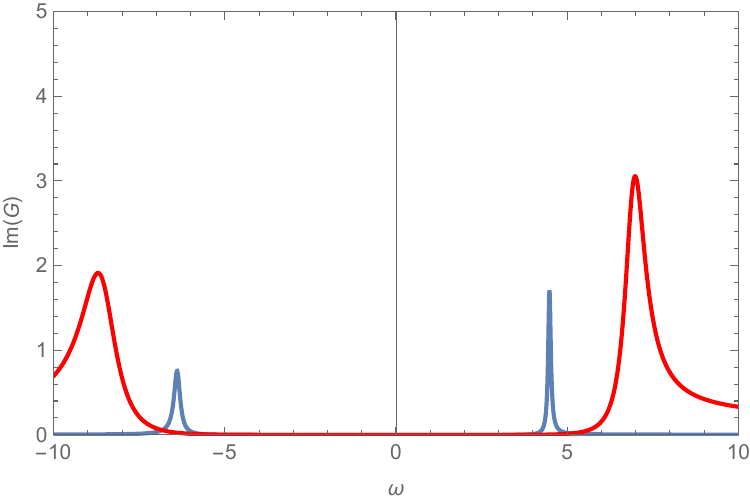}
        \caption{m=-0.49 $\frac{B}{n}=5$}
        \label{}
    \end{subfigure}
    
    \caption{The spectral function for different parameters $m$ and $\frac{B}{n}$ when the momentum relaxation is set to $\beta=0$. The blue, red, and green curves correspond to the Landau levels $l=1, 2, 3$, respectively.}
    \label{fig:1}
\end{figure}

\section{Numerical Results}
\subsection{Spectral functions and Landau level}

\begin{figure}[htbp]
    \centering
    \begin{subfigure}{0.25\textwidth}
        \centering
        \includegraphics[width=\textwidth]{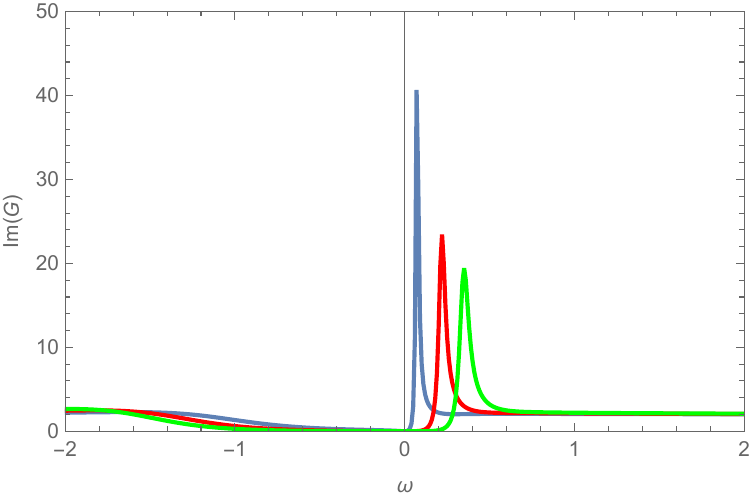}
        \caption{m=0  $\frac{B}{n}=0.2$}
        \label{}
    \end{subfigure}
    \begin{subfigure}{0.25\textwidth}
        \centering
        \includegraphics[width=\textwidth]{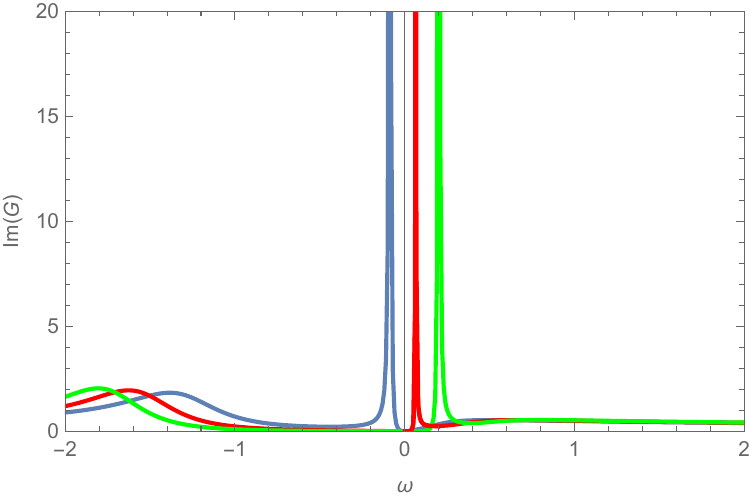}
        \caption{m=-0.25 $\frac{B}{n}=0.2$}
        \label{}
    \end{subfigure}
    \begin{subfigure}{0.25\textwidth}
        \centering
        \includegraphics[width=\textwidth]{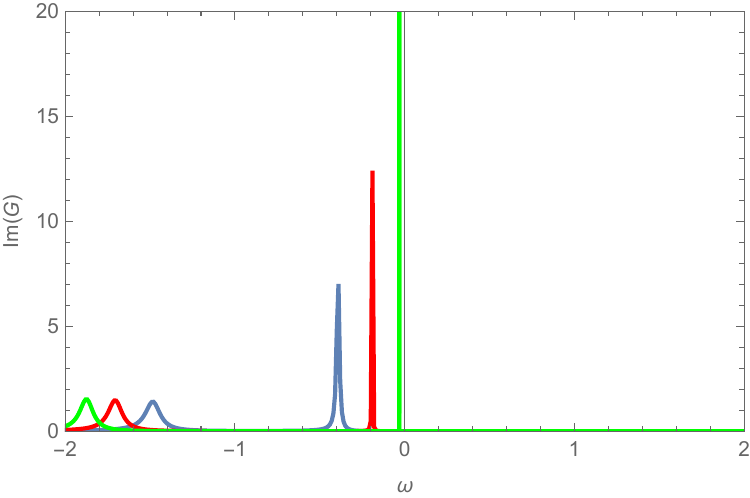}
        \caption{m=-0.49  $\frac{B}{n}=0.2$}
        \label{}
    \end{subfigure}
    
    \begin{subfigure}{0.25\textwidth}
        \centering
        \includegraphics[width=\textwidth]{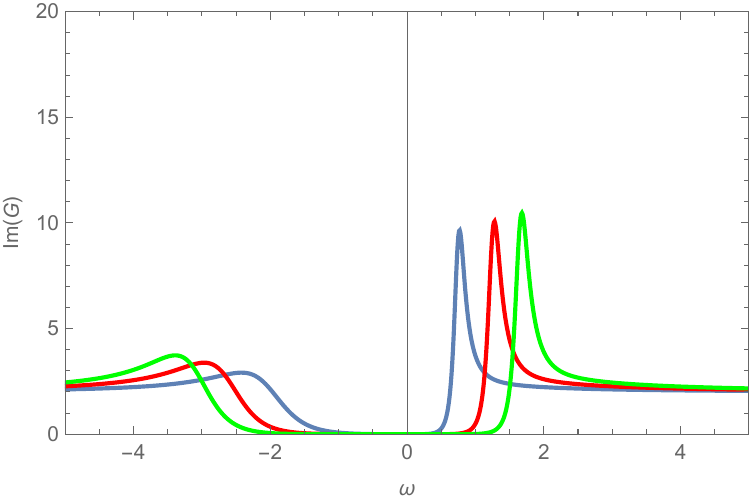}
        \caption{m=0   $\frac{B}{n}=1$}
        \label{}
    \end{subfigure}
    \begin{subfigure}{0.25\textwidth}
        \centering
        \includegraphics[width=\textwidth]{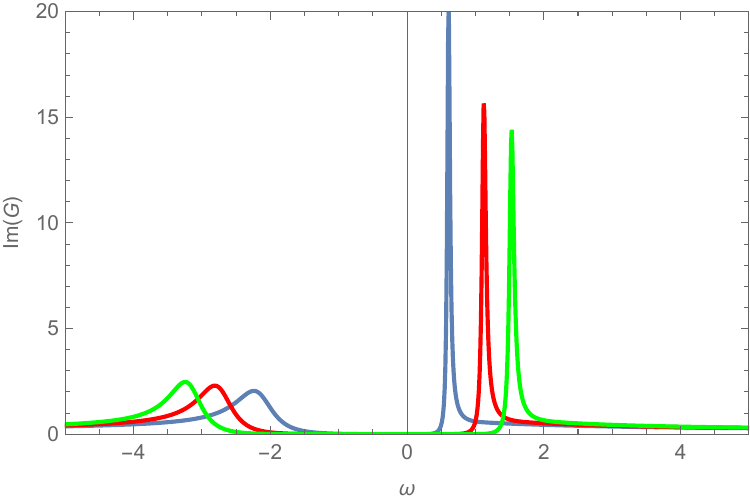}
        \caption{m=-0.25   $\frac{B}{n}=1$}
        \label{}
    \end{subfigure}
    \begin{subfigure}{0.25\textwidth}
        \centering
        \includegraphics[width=\textwidth]{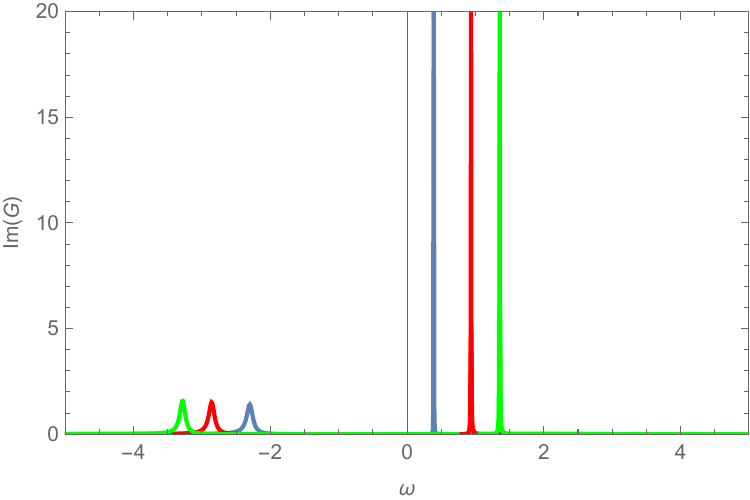}
        \caption{m=-0.49   $\frac{B}{n}=1$}
        \label{}
    \end{subfigure}
    
    \begin{subfigure}{0.25\textwidth}
        \centering
        \includegraphics[width=\textwidth]{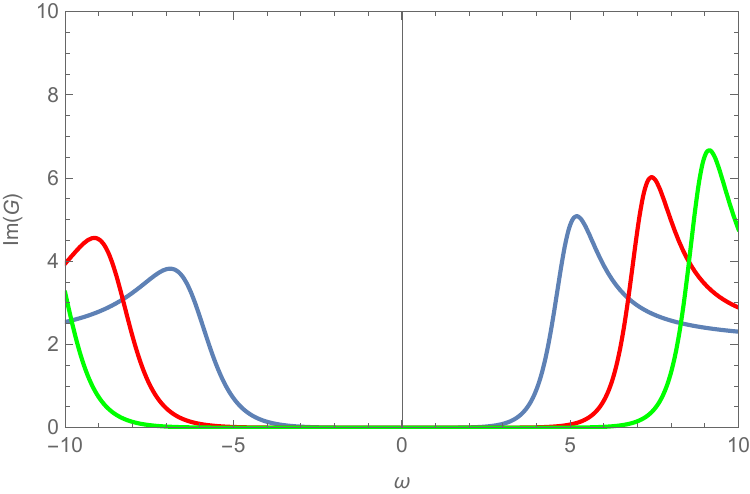}
        \caption{m=0  $\frac{B}{n}=5$}
        \label{}
    \end{subfigure}
    \begin{subfigure}{0.25\textwidth}
        \centering
        \includegraphics[width=\textwidth]{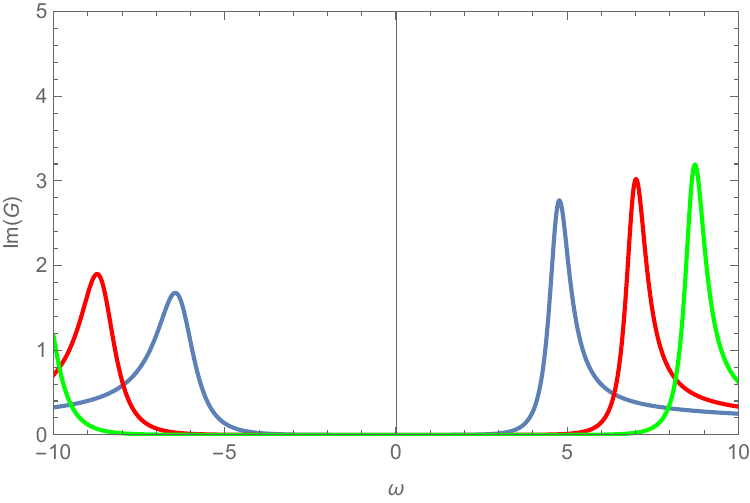}
        \caption{m=-0.25  $\frac{B}{n}=5$}
        \label{}
    \end{subfigure}
    \begin{subfigure}{0.25\textwidth}
        \centering
        \includegraphics[width=\textwidth]{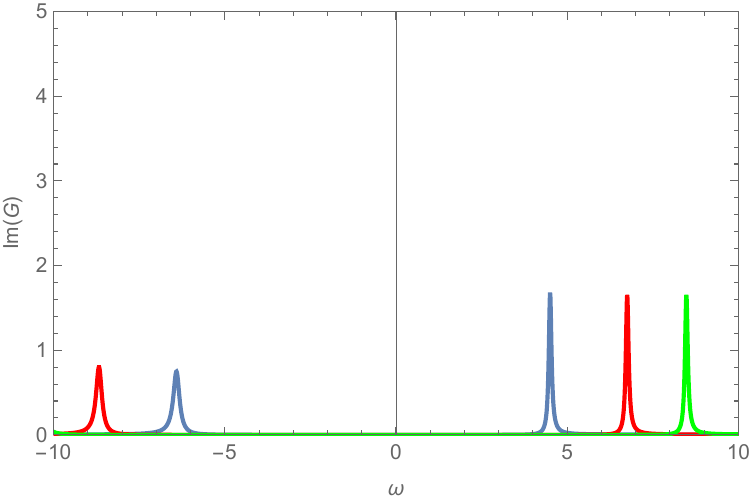}
        \caption{m=-0.49  $\frac{B}{n}=5$}
        \label{}
    \end{subfigure}
    
    \caption{The spectral function for different parameters $m$ and $\frac{B}{n}$ when the momentum relaxation is set to $\beta=0.5$. The blue, red, and green curves correspond to the Landau levels $l=1, 2, 3$, respectively.}
    \label{fig:2}
\end{figure}

\begin{figure}[htbp]
    \centering
    \begin{subfigure}{0.25\textwidth}
        \centering
        \includegraphics[width=\textwidth]{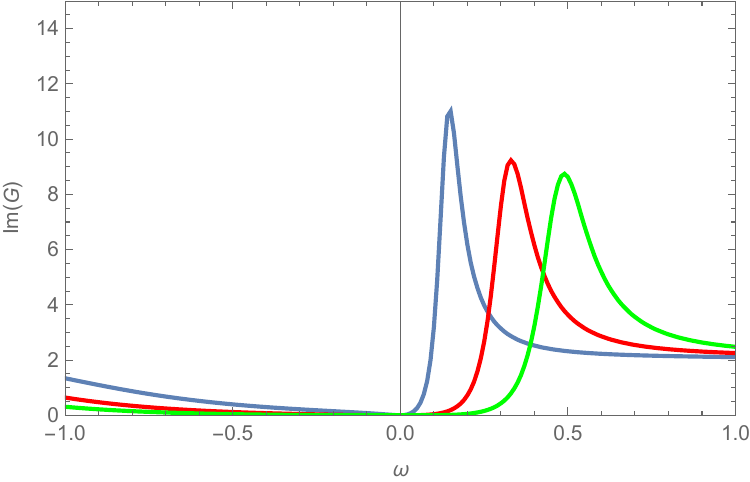}
        \caption{m=0  $\frac{B}{n}=0.2$}
        \label{}
    \end{subfigure}
    \begin{subfigure}{0.25\textwidth}
        \centering
        \includegraphics[width=\textwidth]{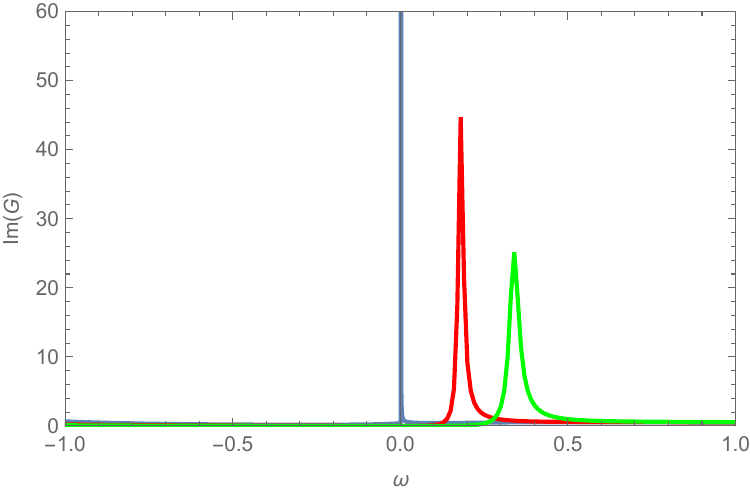}
        \caption{m=-0.25  $\frac{B}{n}=0.2$}
        \label{}
    \end{subfigure}
    \begin{subfigure}{0.25\textwidth}
        \centering
        \includegraphics[width=\textwidth]{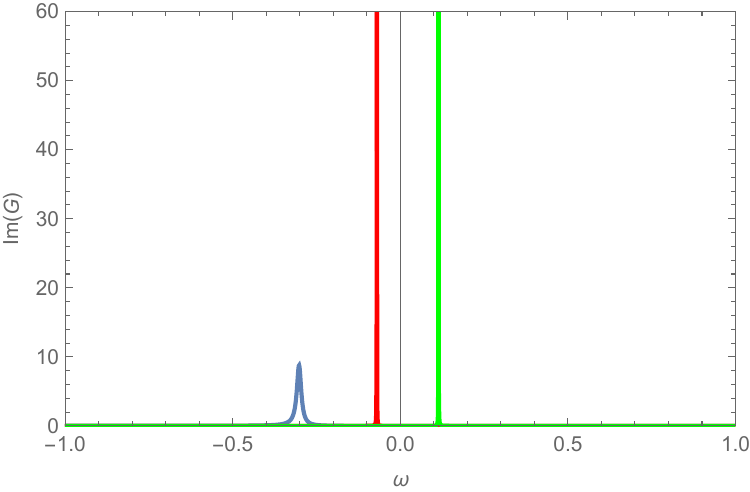}
        \caption{m=-0.49  $\frac{B}{n}=0.2$}
        \label{}
    \end{subfigure}
    
    \begin{subfigure}{0.25\textwidth}
        \centering
        \includegraphics[width=\textwidth]{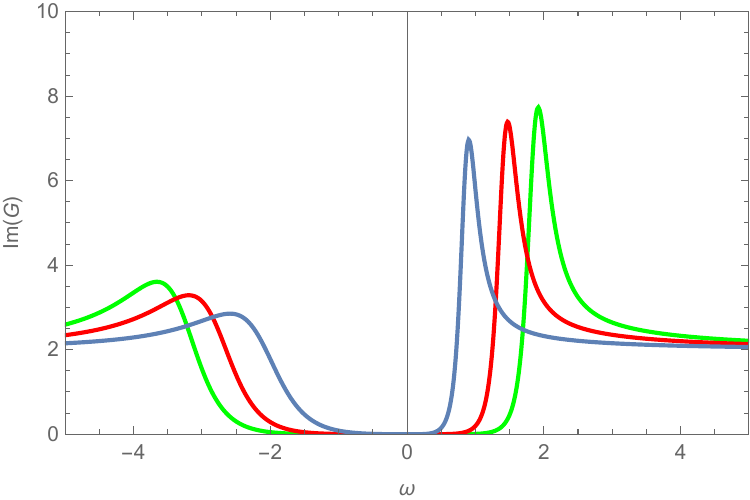}
        \caption{m=0  $\frac{B}{n}=1$}
        \label{}
    \end{subfigure}
    \begin{subfigure}{0.25\textwidth}
        \centering
        \includegraphics[width=\textwidth]{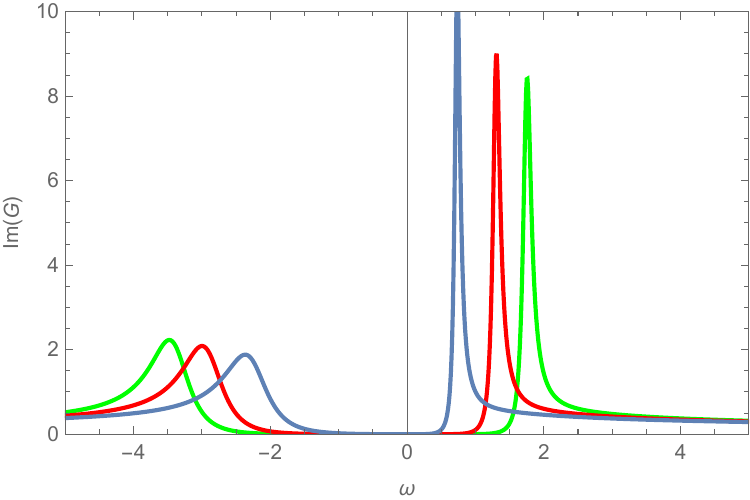}
        \caption{m=-0.25  $\frac{B}{n}=1$}
        \label{}
    \end{subfigure}
    \begin{subfigure}{0.25\textwidth}
        \centering
        \includegraphics[width=\textwidth]{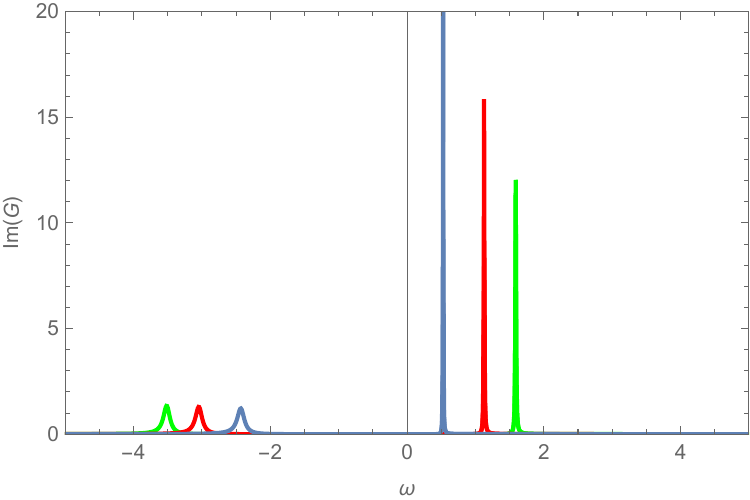}
        \caption{m=-0.49  $\frac{B}{n}=1$}
        \label{}
    \end{subfigure}
    
    \begin{subfigure}{0.25\textwidth}
        \centering
        \includegraphics[width=\textwidth]{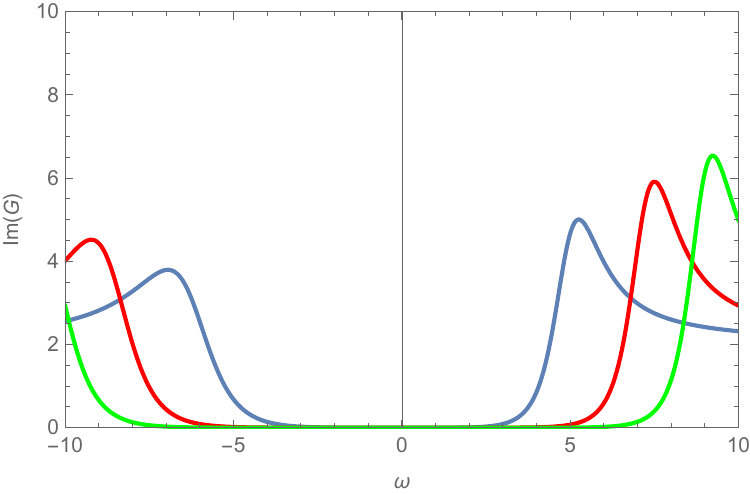}
        \caption{m=0  $\frac{B}{n}=5$}
        \label{}
    \end{subfigure}
    \begin{subfigure}{0.25\textwidth}
        \centering
        \includegraphics[width=\textwidth]{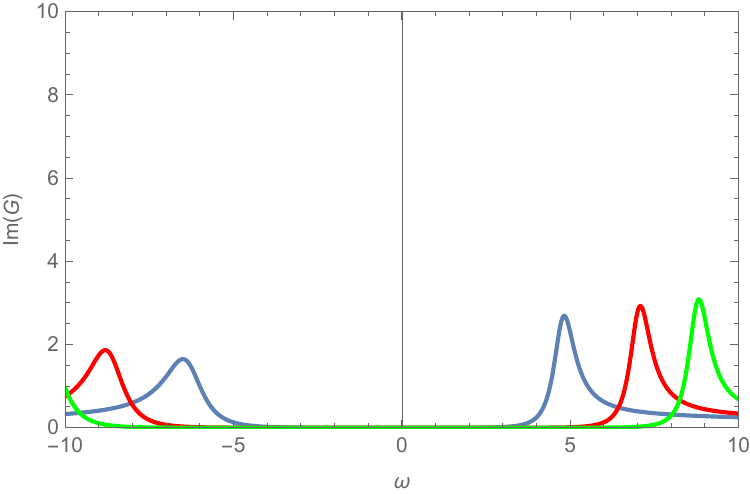}
        \caption{m=-0.25  $\frac{B}{n}=5$}
        \label{}
    \end{subfigure}
    \begin{subfigure}{0.25\textwidth}
        \centering
        \includegraphics[width=\textwidth]{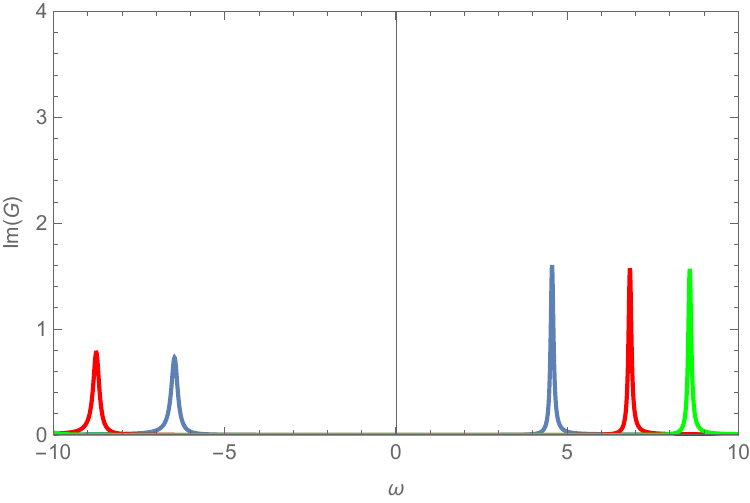}
        \caption{m=-0.49  $\frac{B}{n}=5$}
        \label{}
    \end{subfigure}
    
    \caption{The spectral function for different parameters $m$ and $\frac{B}{n}$ when the momentum relaxation is set to $\beta=1$. The blue, red, and green curves correspond to the Landau levels $l=1, 2, 3$, respectively.}
    \label{fig:3}
\end{figure}

In our study, the gravitational background is described by the momentum relaxation, charge density, and temperature parameters $(\beta, n, T)$. We can utilize \eqref{magnetic} to express the magnetic field $B$ in terms of $\beta, \,n, \,\text{and} \, T$.
The coupled fermion is characterized by mass and charge parameters $(m, q)$. At this point, we work in the zero temperature limit $T=0$. A large value of $q$ leads to the emergence of multiple Fermi surfaces, as noted in \cite{gubser2012analytic}. However, our analysis does not focus on this situation, so we set $q=1$. Because of magnetic effects, the system is divided by many Landau levels. We consider the situation of the Landau levels $l=1,2,3$. These levels are crucial for understanding the transport properties of the system, as they are closest to the Fermi surface. The equation \eqref{flow} can be solved numerically, yielding the spectral functions shown in Figs. \ref{fig:1}, \ref{fig:2}, and \ref{fig:3}.

In these figures, the blue, red, and green curves correspond to Landau levels $l=1,2,3,$ respectively. We first examine the effects of the magnetic field. We set the ratios $\frac{B}{n}=0.2,1,5$ for convenience. This first line of Figs. \ref{fig:1}, \ref{fig:2}, and \ref{fig:3} cosspond to the ratios $\frac{B}{n}=0.2$. It means a weak magnetic field. Obviously, there are three peaks. The peak closest to the $\omega=0$ peak is higher and narrower compared to the others. These peaks represent quasi-particle peaks in the spectral functions. They show that quasiparticle excitation near the Fermi surface is modulated by a weak magnetic field at different Landau energy levels. As the magnetic field strength increases (second row of the figures), the spectral peaks for the Landau levels cross the line $\omega=0$ and become lower and broader. This indicates that these quasiparticle excitations are less stable compared to those in a weak magnetic field. Ultimately, with a further increase in the magnetic field (third row of the figures), the quasiparticle peaks evolve into a nearly symmetric structure, as discussed in \cite{gubankova2011holographic, ge2010analytical}. A large Landau gap at high magnetic fields can account for this behavior, suggesting that our system undergoes metal-insulator phase transitions as the magnetic field strength increases.

Due to the Breitenlohner-Freedman bound, the fermion mass range is restricted to $-\frac{1}{2}<m<\frac{1}{2}$\cite{iqbal2009real}. The first, second, and third columns of these figures correspond to $m=0$,$ −0.25$, and $−0.49$, respectively. As we adjust the mass parameter from 0 to -0.49, these peaks become higher and narrower. Moreover, in a weak magnetic field, the peaks tend to shift across the $\omega=0$ line toward lower $\omega$ in the weak magnetic field, but this shift is suppressed as the magnetic field increases. Adjusting the coupled fermion mass brings the system closer to the behavior of "free fermions". Because the mass reaches the maximum value permitted by the unitarity bound, the conformal dimension of the dual operator reduces to unity \cite{Jeong:2019zab,_ubrovi__2009}. Likewise. a strong magnetic field disrupts the standard quasi-particle peak structure.

By comparing Figs. \ref{fig:1}, \ref{fig:2}, and \ref{fig:3} ($\beta=0,\,0.5,\,\text{and}\,1$), we find that a larger $\beta$ results in lower and broader quasi-particle peaks in these spectral functions, consistent with findings from \cite{Jeong:2019zab}. Additionally, we observe that peaks shift rightward with increasing $\beta$. The broader the peak, the shorter the quasi-particle lifetime. Moving to the right (toward larger $\omega$) corresponds to higher energy. Increased momentum relaxation allows quasi-particles to be excited to higher energy states, potentially leading to instability. This behavior is analogous to the effects observed under a weak magnetic field.

\begin{figure}[htbp]
    \centering
    \begin{subfigure}{0.4\textwidth}
        \centering
        \includegraphics[width=\textwidth]{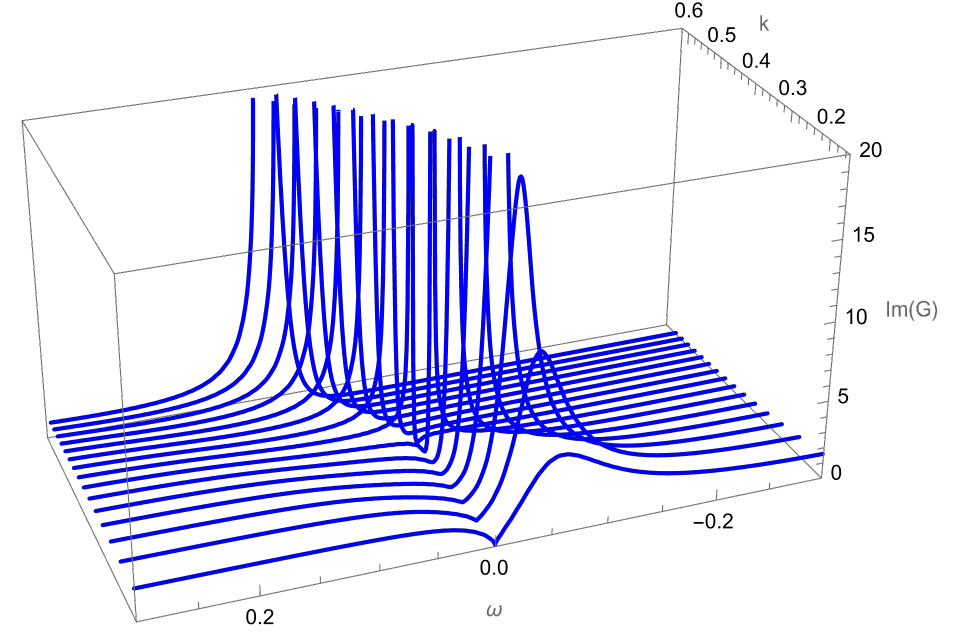}
        \caption{m=0 $\frac{B}{n}=0.02$}
        \label{}
    \end{subfigure}
    \begin{subfigure}{0.4\textwidth}
        \centering
        \includegraphics[width=\textwidth]{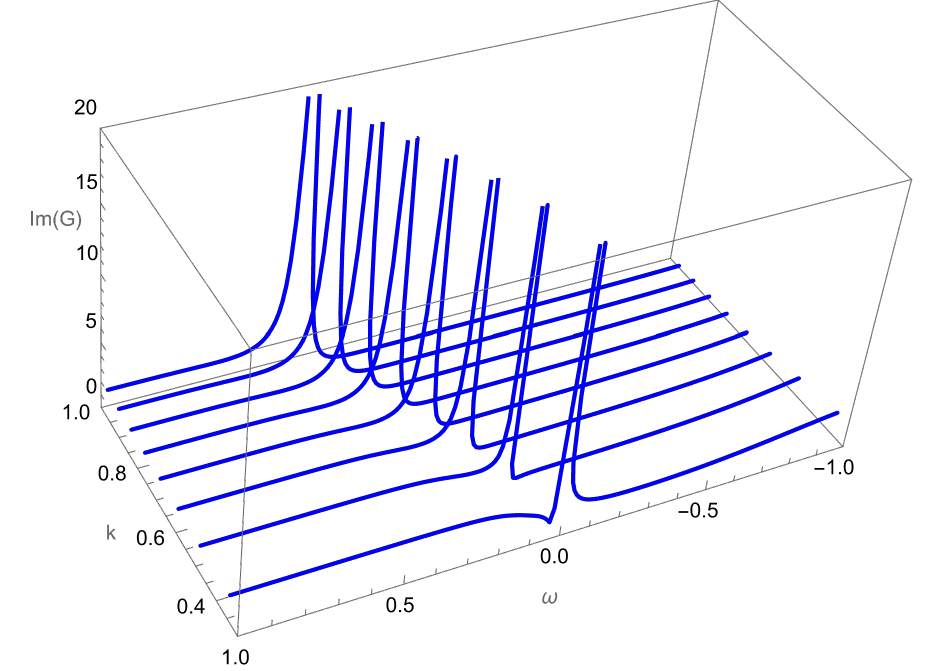}
        \caption{m=0 $\frac{B}{n}=0.1$}
        \label{fig:4b}
    \end{subfigure}
   
    \begin{subfigure}{0.4\textwidth}
        \centering
        \includegraphics[width=\textwidth]{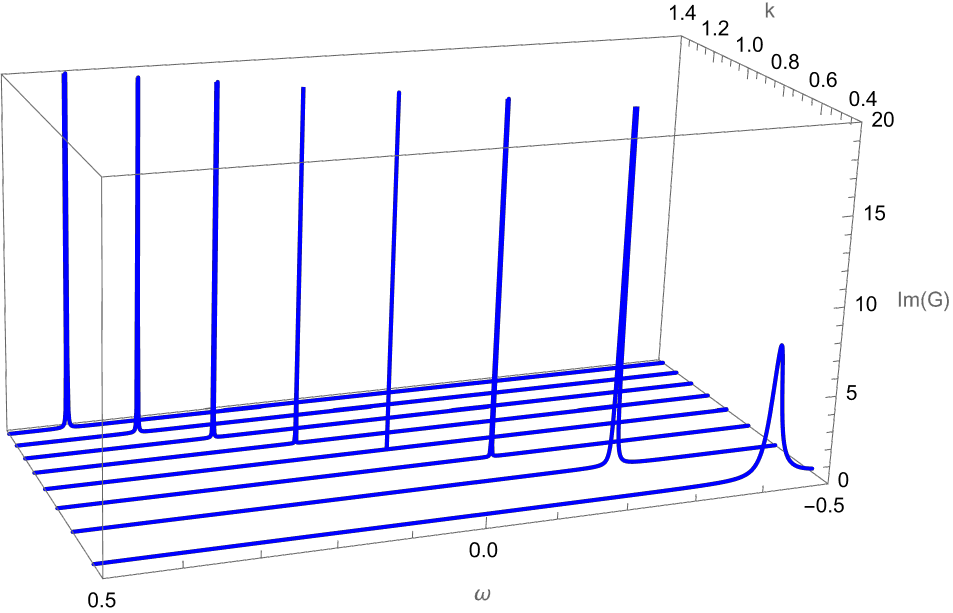}
        \caption{m=-0.49 $\frac{B}{n}=0.2$}
        \label{}
    \end{subfigure}
    \begin{subfigure}{0.4\textwidth}
        \centering
        \includegraphics[width=\textwidth]{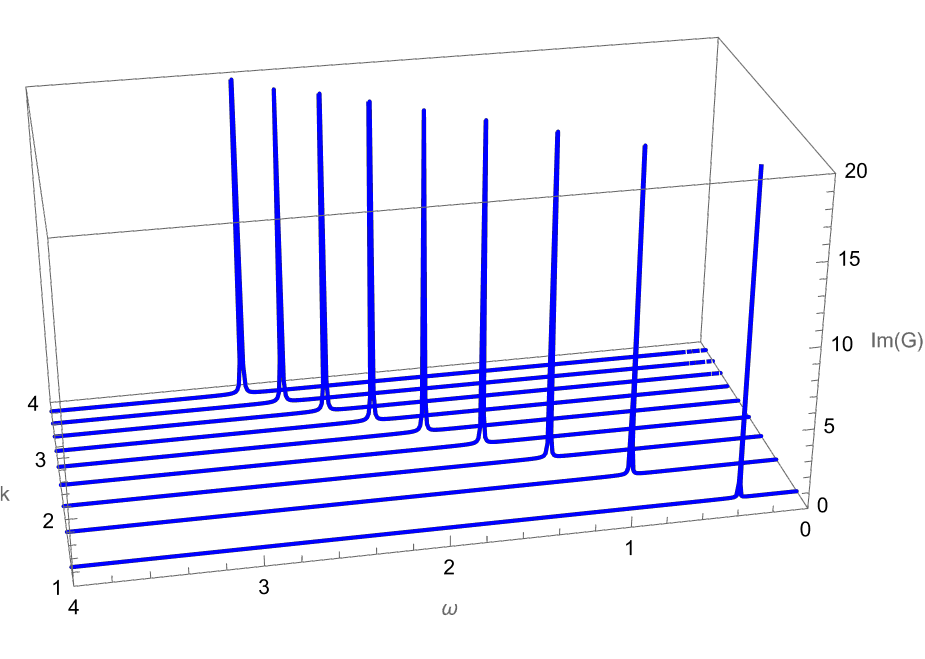}
        \caption{m=-0.49 $\frac{B}{n}=1$}
        \label{}
    \end{subfigure}

    \caption{The imaginary part of the Green's function, $\text{Im}\, G(k, \omega)$, is presented as a function of momentum and frequency for various values of the parameter $\frac{B}{n}$, with the momentum relaxation set to $\beta = 0$. The different blue curves correspond to distinct Landau levels.}
    \label{fig:4}
\end{figure}

\begin{figure}[htbp]
    \centering
    \begin{subfigure}{0.4\textwidth}
        \centering
        \includegraphics[width=\textwidth]{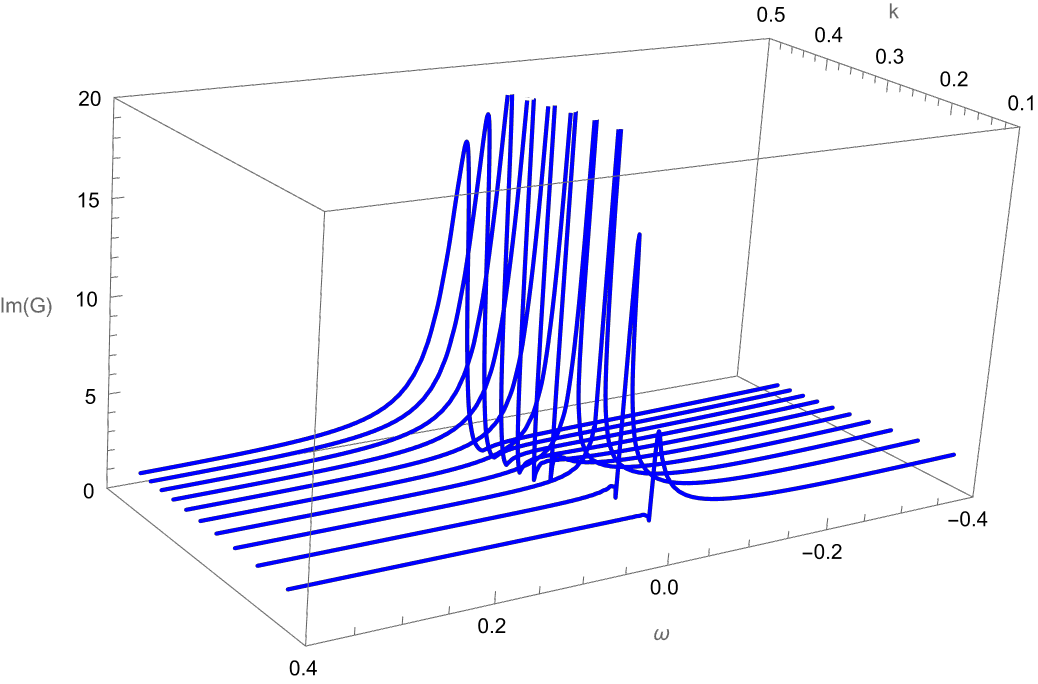}
        \caption{m=0 $\frac{B}{n}=0.01$}
        \label{}
    \end{subfigure}
    \begin{subfigure}{0.4\textwidth}
        \centering
        \includegraphics[width=\textwidth]{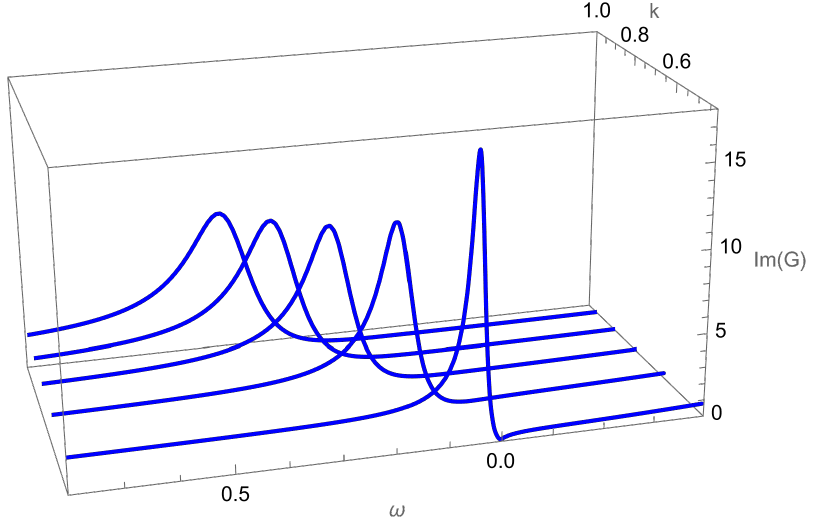}
        \caption{m=0 $\frac{B}{n}=0.1$}
        \label{}
    \end{subfigure}

    \begin{subfigure}{0.4\textwidth}
        \centering
        \includegraphics[width=\textwidth]{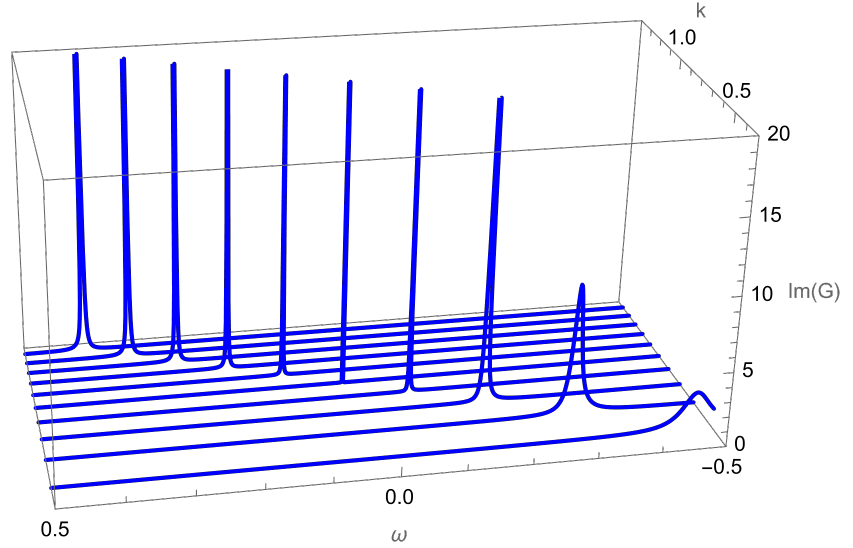}
        \caption{m=-0.49 $\frac{B}{n}=0.1$}
        \label{}
    \end{subfigure}
    \begin{subfigure}{0.4\textwidth}
        \centering
        \includegraphics[width=\textwidth]{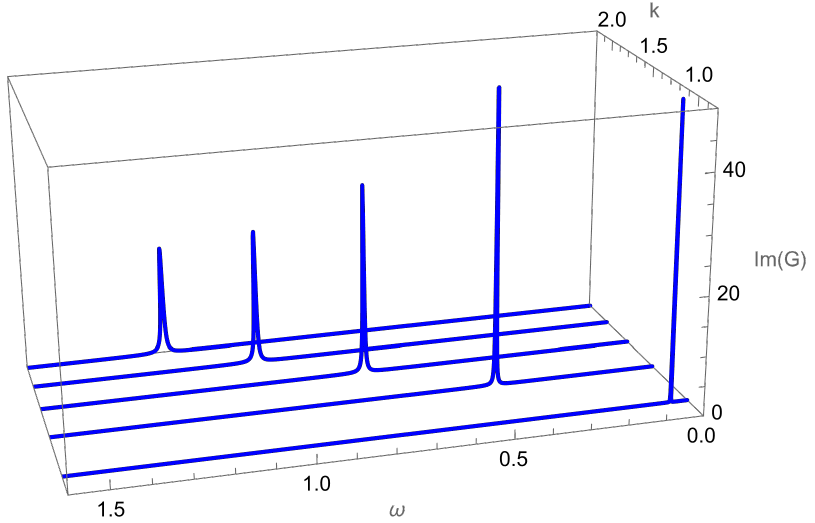}
        \caption{m=-0.49 $\frac{B}{n}=0.5$}
        \label{fig}
    \end{subfigure}

    \caption{The imaginary part of the Green's function, $\text{Im}\, G(k, \omega)$, is presented as a function of momentum and frequency for various values of the parameter $\frac{B}{n}$, with the momentum relaxation set to $\beta = 1$. The different blue curves correspond to distinct Landau levels.}
    \label{fig:5}
\end{figure}

To provide a clearer picture of the spectral function and Fermi momentum, we present Figs. \ref{fig:4} and \ref{fig:5}. Here, we care about the weak magnetic field regime for choosing some small value of $\frac{B}{n}$. These figures illustrate several spectral functions in the momentum-energy space. They are limited by some Landau levels in the momentum direction, with the lowest being $l = 1$. As shown in the first column of Figs.\ref{fig:4} and Figs.\ref{fig:5}, quasi-particle peaks at some Landau levels lie below Fermi momentum. As the magnetic field intensifies (the second column of these figures), more Landau levels cross the Fermi momentum, while the intervals between successive Landau levels increase. Ultimately, the lowest Landau level would extend beyond the Fermi momentum. Quasi particles near these Fermi surfaces, which are confined by Landau levels, have higher energies as the magnetic field increases. By comparing Fig.\ref{fig:4} to Fig.\ref{fig:5}, increased momentum relaxation leads to a rapid decline in the peaks of the spectral functions near the Fermi momentum. This means that the larger momentum relaxation makes the quasi-particle excitation near the Fermi surface more unstable.

When $m=-0.49$, the system exhibits a clear quasi-particle picture and a linear dispersion relationship from Fig.\ref{fig:4} and Fig.\ref{fig:5}. As we adjust the mass to zero, these quasi-particle peaks become lower and broader, and the dispersion relation transitions to a non-linear form. These phenomena again support our previous view that when the mass of the coupled fermions approaches $-0.49$, the system becomes much like some "free fermions".

In this section, we present the spectral function numerically for various parameter sets. As the magnetic field strength or the momentum relaxation increases, quasi-particles near the Fermi surface become more unstable with higher energies. The difference is that increasing the magnetic field makes the quasi-particle picture entirely disappear, evolving into a near-symmetric structure. This indicates a metal-insulator phase transition. Due to the increase of the magnetic field, the Landau level moves in the higher energy direction and sweeps through the Fermi surface. The minimum allowable quasi-particle energy of the system could be higher than the Fermi energy, and the spacing between Landau levels will be increased. All these phenomena align with established condensed matter theory \cite{kittel2021introduction}, which indicates that increased magnetic fields elevate the total electron energy. This behavior is characteristic of quantum oscillations, resulting in notable phenomena such as the de Haas-van Alphen effect and the Shubnikov-de Haas effect.

\subsection{Fermi momentum and scaling exponents}
We investigate how the system responds to the effects of magnetic fields and momentum relaxation, setting the mass $m=0$ and considering the zero-temperature limit, $T=0$. The Fermi surface is a crucial quantity for any fermionic system. As discussed in the previous section, strong magnetic effects disrupt the structure of the spectral function, rendering the definition of the Fermi surface ambiguous in high magnetic fields. To explore this further, we consider the effective momentum $k=\lambda$ by numerically solving the Dirac equation \eqref{flow}, while neglecting the discrete momentum effects due to the magnetic field.

\begin{figure}[htbp]
    \centering    \includegraphics[width=0.45\linewidth]{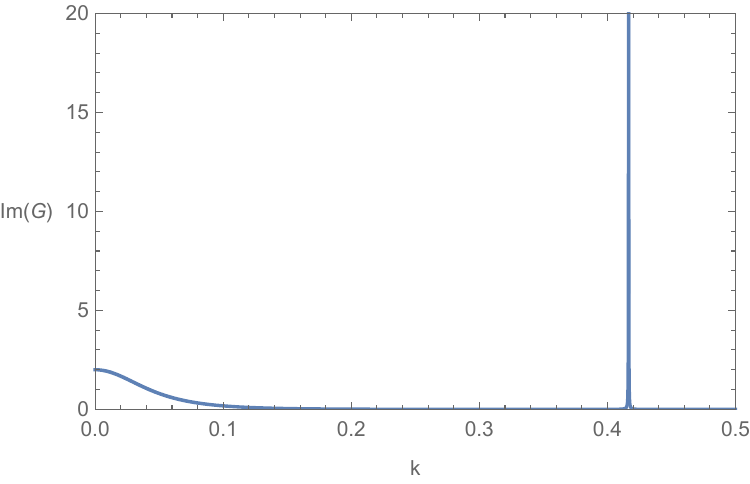}
    \caption{Plots of $\text{Im}\, G (k)$ as a function of $k$ at $\omega=-10^{-8} \,(\beta=0 \, \text{and}\, B=0 )$. A
 sharp peak in $\text{Im}\, G (k)$ is visible at $k_F\approx0.4165$.}
    \label{fig:}
\end{figure}

In Fig.\ref{fig:}, we present the imaginary part of the Green's function, $\text{Im} 
\, G(k)$, calculated by setting $\omega = -10^{-8}$ and numerically solving the Dirac equation in the absence of both momentum relaxation and the magnetic field. A sharp and prominent peak is observed, indicating the presence of a Fermi surface in the dual field theory. The position of this peak corresponds to the Fermi momentum, $k_F \approx 0.4165$. Next, we introduce the magnetic field and momentum relaxation into the gravitational background to explore the resulting phenomena in the dual field theory.

\begin{figure}[htbp]
    \centering
    \begin{subfigure}{0.45\textwidth}
        \centering
        \includegraphics[width=\textwidth]{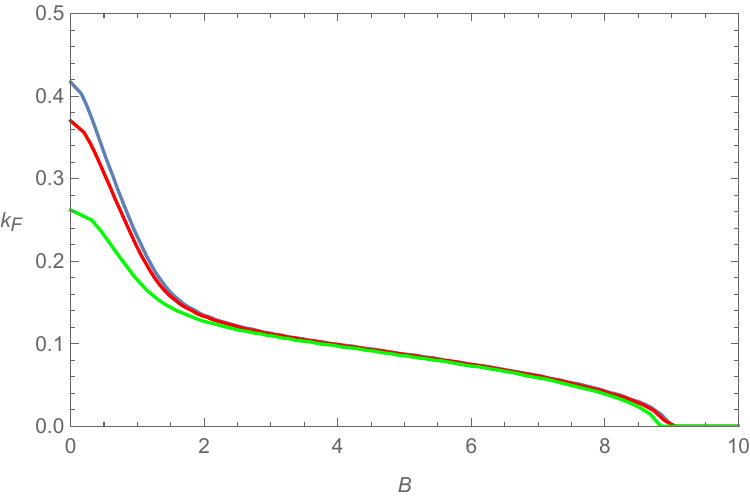}
        \caption{}
        \label{fig:6a}
    \end{subfigure}
    \begin{subfigure}{0.45\textwidth}
        \centering
        \includegraphics[width=\textwidth]{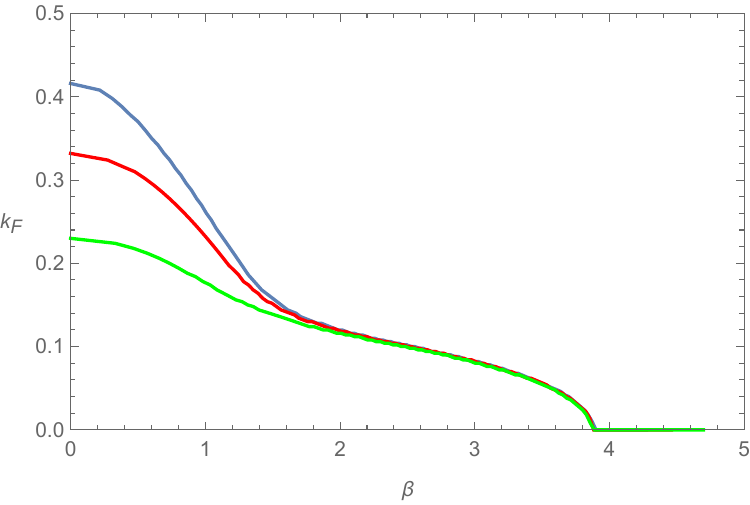}
        \caption{}
        \label{fig:6b}
    \end{subfigure}
    \caption{(a) The Fermi momentum $k_F$ as a function of magnetic field $B$. The blue, red, and green colors correspond to the momentum relaxation $\beta=0,0.5,1$; (b) The Fermi momentum $k_F$ as a function of momentum relaxation $\beta$. The blue, red, and green colors correspond to the magnetic field $B=0,0.5,1$.}
    \label{fig:6}
\end{figure}

\begin{figure}[htbp]
    \centering
    \begin{subfigure}{0.45\textwidth}
        \centering
        \includegraphics[width=\textwidth]{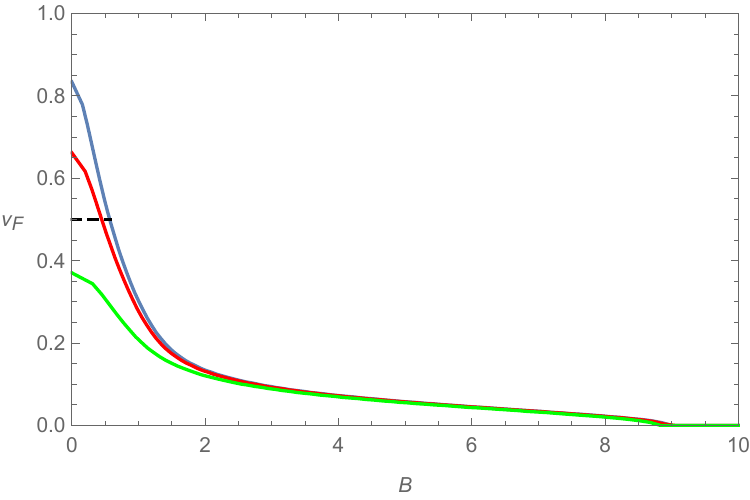}
        \caption{}
        \label{fig:7a}
    \end{subfigure}
    \begin{subfigure}{0.45\textwidth}
        \centering
    \includegraphics[width=\textwidth]{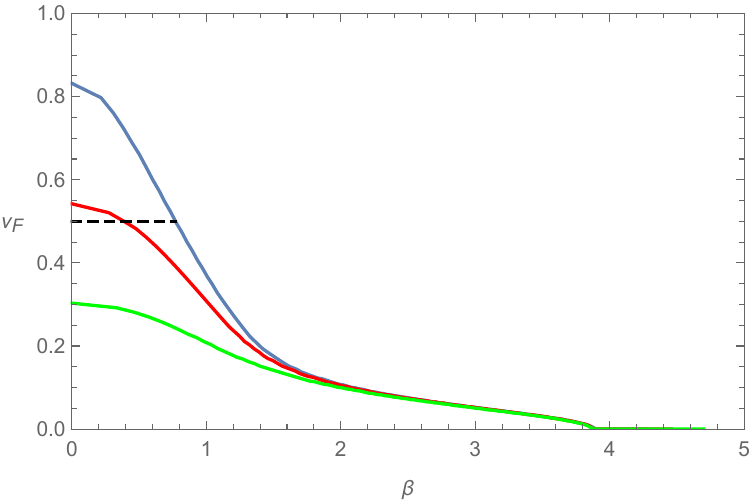}
        \caption{}
        \label{fig:7b}
    \end{subfigure}
    \caption{(a) The scaling exponent $\nu_F$ as a function of magnetic field $B$. The blue, red, and green colors correspond to the momentum relaxation $\beta=0,0.5,1$;(b) The scaling exponent $\nu_F$ as a function of momentum relaxation $\beta$. The blue, red, and green colors correspond to the magnetic field $B=0,0.5,1$}
    \label{fig:7}
\end{figure}

Fig.\ref{fig:6a} and Fig.\ref{fig:7a} show the magnetic field effects on this system. In Fig.\ref{fig:6a}, $k_F$ is the effective Fermi momentum. We can obtain this by numerically solving the Dirac equation \eqref{flow} with $\omega=-10^{-8}$. The momentum relaxations, $ \beta = 0, \,0.5,\, 1$ from top to bottom, distinguish three curves.  The effective Fermi momentum decreases to zero with increasing magnetic field. Our model exhibits a distinct trend compared to that in \cite{gubankova2011holographic}, although both show that the Fermi surface contracts and eventually vanishes under strong magnetic fields. The different colors represent varying degrees of momentum relaxation, indicating that increased momentum relaxation leads to further shrinkage of the Fermi surface. Next, we focus on another significant quantity, the scaling exponent at the fermi momentum, $\nu_F = \nu_{k_F}$, which characterizes the boundary of the non-Fermi liquid regime \cite{faulkner2011emergent}.

In the last section, we have obtained the scaling exponent $\nu$ \eqref{nu}. By substituting the numerical $k_F$ and the selected parameters ($ \beta = 0, \,0.5,\, 1$ from top to bottom as shown in Fig.\ref{fig:7}) into \eqref{nu}
\begin{align}
    \nu_F=  2 k_F \sqrt{\frac{2}{6n^2-\beta^2}}.
\end{align}
We can obtain the function of $\nu_F$ with respect to the magnetic field or momentum relaxation as shown in Fig.\ref{fig:7}. The black dashed line indicates $\nu_F=\frac{1}{2}$. A value of $\nu_F$ greater than $\frac{1}{2}$ indicates that the system behaves as a Fermi liquid, while $\nu_F$ signifies a marginal Fermi liquid. If $\nu_F$ is less than $\frac{1}{2}$, the system is classified as a non-Fermi liquid. Increasing the magnetic field leads to a phase transition from the Fermi liquid regime to the non-Fermi liquid regime. Similarly, an increase in momentum relaxation also results in a decrease in $\nu_F$, facilitating the transition from the Fermi liquid regime to the non-Fermi liquid regime.

Fig.\ref{fig:6b} and Fig.\ref{fig:7b} mainly illustrate the effects of momentum relaxation on the system. The blue, red, and green colors represent different magnetic field strengths, $B = 0, \,0.5,\, 1$ from top to bottom. As momentum relaxation increases, both the effective Fermi momentum $k_F$ and the scaling exponent $\nu_F$ decrease, eventually approaching zero. The former indicates a shrinking of the Fermi surface, which eventually vanishes, while the latter signals a transition from the Fermi liquid regime to the non-Fermi liquid regime. In \cite{Fang:2015dia}, the authors utilize an Einstein-Maxwell gravity model with a massless scalar field, where momentum relaxation is interpreted as arising from an effective impurity. As the momentum relaxation parameter governing this effect increases, they observe a phase transition consistent with the results we have obtained.

\subsection{Dispersion relations}
We know, that the dispersion relations of low-energy quasiparticles near the Fermi surface in a Fermi liquid are expected to be linear. In the previous subsection, we identified a phase transition from the Fermi liquid regime to a non-Fermi liquid regime, characterized by varying magnetic fields and momentum relaxation. Therefore, it is very important to analyze how the dispersion relations evolve under the influence of these magnetic fields and momentum relaxation effects. 

\begin{figure}[htbp]
    \centering    \includegraphics[width=0.45\linewidth]{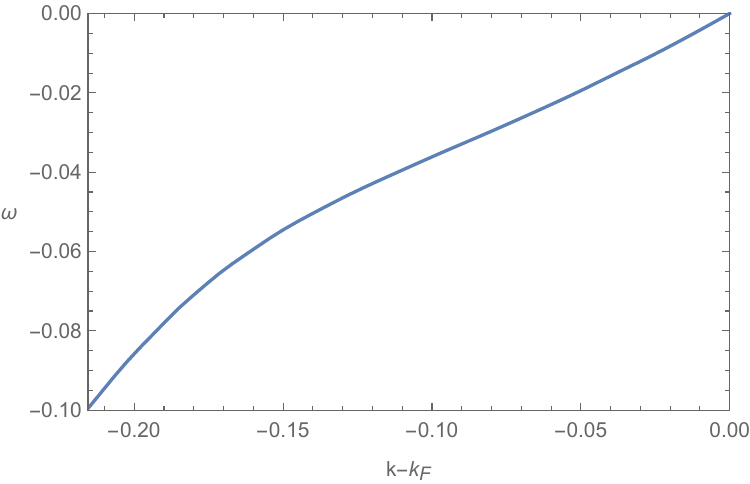}
    \caption{The dispersion relation $\omega (k-k_F)$ near the Fermi surface at the magnetic field $B=0$ and the momentum relaxation $\beta=0$.}
    \label{fig:8}
\end{figure}
We analyze the dispersion relation of quasi-particles within the frequency range $\omega \in (-0.1, 0)$, focusing on their behavior beneath the Fermi surface. As illustrated in Fig.\ref{fig:8}, the dispersion relation for the low-frequency is linear when both momentum relaxation and the magnetic field are absent. Under these conditions, our model simplifies to the original Gubser-Rocha black hole, whose metric does not include terms related to momentum relaxation or magnetic field. This result aligns with the findings for the original Gubser-Rocha black hole, where the dispersion relation of holographic fermions near the Fermi surface was found to be precisely linear \cite{li2012holographic,wu2011some}. Furthermore, as seen in Fig.\ref{fig:7}, when both the magnetic field and momentum relaxation parameters are absent, the scaling exponent $\nu_F$ exceeds $\frac{1}{2}$. This indicates that the system behaves as a Fermi liquid, consistent with the characteristic linear dispersion relation. The system's dispersion relation naturally exhibits nonlinear behavior as the frequency increases.

\begin{figure}[htbp]
    \centering
    \begin{subfigure}{0.45\textwidth}
        \centering
        \includegraphics[width=\textwidth]{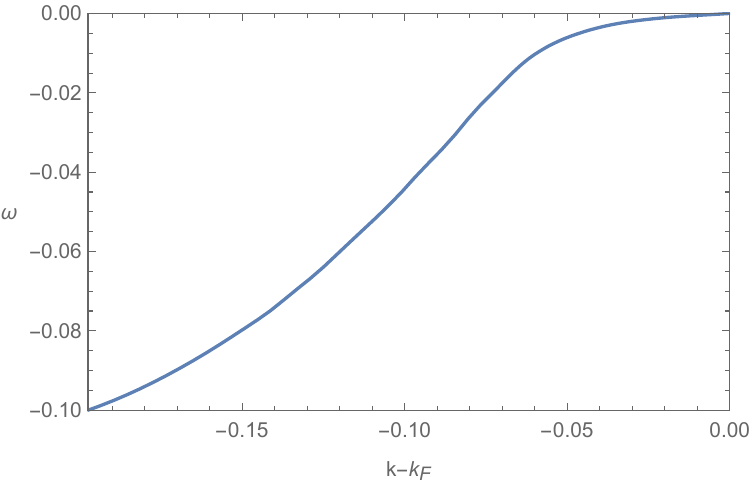}
        \caption{}
        \label{fig:9a}
    \end{subfigure}
    \begin{subfigure}{0.45\textwidth}
        \centering
        \includegraphics[width=\textwidth]{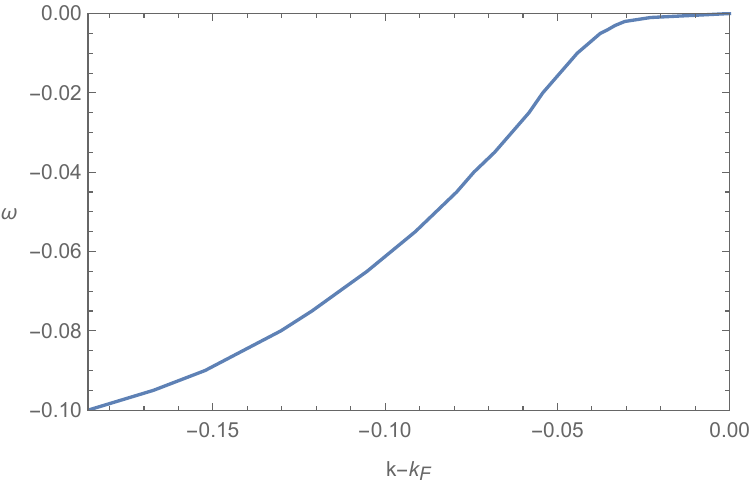}
        \caption{}
        \label{fig:9b}
    \end{subfigure}
    \caption{((a)) The dispersion relation $\omega  (k-k_F)$ near the Fermi surface at the magnetic field $B=0$ and the momentum relaxation $\beta=1$; ((b)) The dispersion relation $\omega  (k-k_F)$ near the Fermi surface at the magnetic field $B=1$ and the momentum relaxation $\beta=0$.}
    \label{fig:9}
\end{figure}
The dispersion relation for the low-frequency becomes non-linear when momentum relaxation is introduced with $\beta=1$, as shown in Fig.\ref{fig:9a}. Compared to the case without a magnetic field and momentum dissipation (Fig. \ref{fig:8}), the low-frequency quasiparticle excitations are shifted further away from the Fermi surface. This indicates that the system transitions into the non-Fermi liquid regime. Furthermore, this observation is consistent with the prediction of a non-Fermi liquid behavior based on the scaling exponent shown in Fig.\ref{fig:7}. Specifically, when the momentum relaxation parameter is set to $\beta=1$ and the magnetic field parameter is zero, the scaling exponent of the system is less than $\frac{1}{2}$, a hallmark characteristic of a non-Fermi liquid.

Next, we examine the effect of a magnetic field with $B=1$, as shown in Fig.\ref{fig:9b}. In this scenario, the scaling exponent, as indicated in Fig.\ref{fig:7}, is less than $\frac{1}{2}$, suggesting that the system resides in the non-Fermi liquid regime. Consistently, Fig.\ref{fig:9b} shows that introducing a significant magnetic field leads to non-linear behavior in the dispersion relation for the low-frequency. Compared with Fig.\ref{fig:9a}, the degree of non-linearity varies depending on the specific values of the applied magnetic field or momentum relaxation, highlighting their distinct contributions to the system's behavior.

In this section, our model demonstrates the phase transition from the Fermi liquid regime to the non-Fermi liquid regime as both the magnetic field and momentum relaxation increase, based on the evaluation of the scaling exponent. A large Fermi surface and linear dispersion relations characterize the Fermi liquid regime. In contrast, the non-Fermi liquid regime exhibits non-linear dispersion relations, with the Fermi surface progressively shrinking, and potentially disappearing altogether.

\section{Magneto-Scattering Rate}
In this section, we focus on an important quantity in condensed matter physics: the scattering rate. Previous studies \cite{Mauri:2024wgc,faulkner2011emergent} indicate that the quasi-particle scattering rate $\gamma$ is given by the imaginary part of the self-energy  $\Sigma''$, which is related to the imaginary part of the infrared (IR) Green function $\rm{Im}(\mathcal{G}_{IR})$, 
\begin{align}
    \gamma \propto \Sigma'' \propto \rm{Im}(\mathcal{G}_{IR}).
\end{align}
At low energy and low temperature, the imaginary part of the full solution is accurately described by the IR results \cite{Mauri:2024wgc}.  However, the effects of the magnetic field have not been considered in these studies. We aim to investigate the dependence of the scattering rate on the magnetic field.

\begin{figure}[htbp]
    \centering    \includegraphics[width=0.45\linewidth]{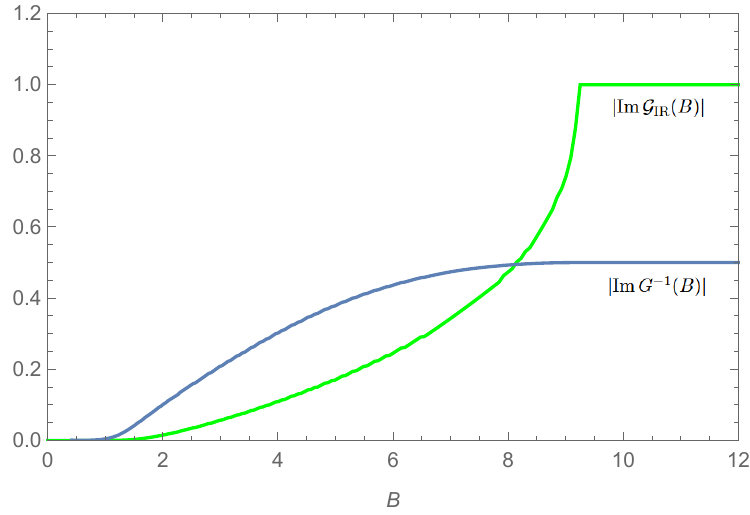}
    \caption{The green curve: $|\text{Im} \,\mathcal{G}_{\text{IR}}|$ as a function of magnetic field $B$. The blue curve: $|\text{Im} \, G^{-1}(B)|$ as a function of magnetic field $B$.}
    \label{fig:11}
\end{figure}

Using the ultraviolet (UV) Green's function \eqref{G^-1} and the infrared (IR) result \eqref{gamma} at zero temperature, we set the momentum relaxation parameter $\beta = 0$ and plot $|\rm{Im}(G^{-1})|$ and $|\rm{Im}(\mathcal{G}_{\rm{IR}})|$ in Fig.\ref{fig:11}. In the Fermi liquid regime, both quantities remain small. However, as the magnetic field increases, their behaviors diverge significantly in the non-Fermi liquid regime. Eventually, the scattering rate reaches a maximal value, as shown in the plateau region of Fig.\ref{fig:11}.

To analyze IR Green's function \eqref{IR Green}, we note that the flat region of the green curve corresponds to the disappearance of the Fermi surface. Substituting $\nu_F = 0$ into \eqref{gamma}, we find $|\text{Im},\mathcal{G}_{\text{IR}}(B)| = 1$. For the blue curve, which represents the reciprocal of the UV Green's function, we also identify a maximal value. Both results indicate that the system ultimately transitions to an insulating state. The difference lies in the nature of the transition: the UV results exhibit a smooth evolution.

Notably, regardless of whether we analyze the UV Green's function \eqref{G^-1} or the IR Green's function \eqref{IR Green}, both approaches reveal qualitatively similar behavior for the scattering rate. This consistency highlights the system's evolution from a Fermi liquid to a non-Fermi liquid, and eventually to an insulating state. 

\section{Discussion and Conclusion}
We have investigated the fermionic response of the dyonic Gubser-Rocha model at zero temperature. The model's near-horizon geometry exhibits a conformal $\rm{AdS_2}$ structure. Solving the near-horizon Dirac equation, we obtained the scaling exponent \eqref{nu}, revealing that both momentum relaxation and the magnetic field significantly affect the IR physics and low-energy excitations.

We explored the spectral function under various parameter regimes through numerical solutions of the bulk Dirac equation. A strong magnetic field induces substantial changes in the spectral structure, transitioning from a quasi-particle picture to a near-symmetric form, as demonstrated in Figs.\ref{fig:1}, \ref{fig:2}, and \ref{fig:3}. These figures reveal the emergence of the Landau gap, signaling a metal-insulator transition. At weak magnetic fields, the quasi-particle picture persists, but the system is divided by Landau levels, as seen in Figs.\ref{fig:4} and \ref{fig:5}. As the magnetic field increases, the separation between adjacent Landau levels widens, and these levels shift toward higher momentum. In the limit of a very strong magnetic field, all Landau levels exceed the Fermi surface, indicating a threshold magnetic field where no states remain below the Fermi surface.

We also analyzed key quantities, including the Fermi momentum, and scaling exponent (Figs.\ref{fig:6} and \ref{fig:7}). As the magnetic field increases, disregarding its discrete effects, we observe a progressive Fermi surface shrink and a decrease in the scaling exponent. This behavior indicates a transition from a Fermi liquid to a non-Fermi liquid, ultimately culminating in an insulating state. The quasi-particle dispersion near the Fermi surface, shown in Figs.\ref{fig:8} and \ref{fig:9}, correspond to different magnetic field configurations and momentum relaxation. In Fig.\ref{fig:8}, both the magnetic field and momentum relaxation are set to zero ($B = \beta = 0$), corresponding to a Fermi liquid state characterized by linear dispersion in the low-energy region. In contrast, Figs.\ref{fig:9a} and \ref{fig:9b} show the dispersion for configurations with either a nonzero magnetic field or momentum relaxation. In these cases, the system transitions into the non-Fermi liquid regime, where the dispersion becomes more nonlinear compared to the scenario without magnetic field and momentum relaxation.
 
We further examined the dependence of the scattering rate on the magnetic field, using both IR results and the UV Green's function. As shown in Fig.\ref{fig:11}, the scattering rates are small in the Fermi liquid regime but increase and diverge in the non-Fermi liquid regime. Ultimately, they reach a maximum value as the Fermi surface vanishes, marking the transition to an insulating state. And the UV results exhibit a smoother evolution compared to the IR results.

In the dyonic Gubser-Rocha model, when both the magnetic field and momentum relaxation are set to zero, the original Gubser-Rocha model is recovered, retaining the same scaling exponent and linear dispersion relation at low energy. However, the introduction of the magnetic field and momentum relaxation extends the scaling exponent and alters the infrared physics. In the dual $(2+1)$ dimensional field theory, we observed a phase transition where the system evolves from a Fermi liquid to a non-Fermi liquid, and ultimately to an insulating state as either the magnetic field or momentum relaxation increases. This "metal"-insulator transition has been experimentally observed in some low-dimensional systems and semiconductors \cite{PhysRevLett.72.709,PhysRevLett.79.1543,PhysRevB.104.045138,PhysRevLett.57.1056}. The spectral structure undergoes a notable transformation, with the linear dispersion relation transitioning into a nonlinear form at low energies as the magnetic field and momentum relaxation are varied.

We conclude that the holographic framework is a powerful tool for studying strongly coupled fermionic systems. By directly analyzing the IR near-horizon physics, we gain insight into low-energy excitations without relying on the quasi-particle picture, avoiding the complexities of traditional field theory analyses. We hope that this approach contributes to the understanding of high-temperature superconductivity and strange metals.

\section*{Acknowledgement}
We would like to thank ZhaoJie Xu, LiQing Fang and WenBin Li for their helpful discussions. The study was partially supported by NSFC, China (grant No.12275166 and  No.12311540141). This work is also supported by NRF of Korea with grant No.NRF-2021R1A2B5B02002603, RS-2023-00218998 and NRF 2022H1D3A3A01077468.

\bibliographystyle{unsrt}  
\bibliography{references}  

\begin{thebibliography}{10}

\bibitem{Landau:1958joj}
Lev~Davidovich Landau.
\newblock {On the Theory of the Fermi Liquid}.
\newblock {\em J. Exp. Theor. Phys.}, 35, 1958.

\bibitem{bednorz1986possible}
J~George Bednorz and K~Alex M{\"u}ller.
\newblock {Possible high $T_c$ superconductivity in the Ba-La-Cu-O system}.
\newblock {\em {Z. Physik B - Condensed Matter}}, 64(2):189--193, 1986.

\bibitem{Maldacena:1997re}
Juan~Martin Maldacena.
\newblock {The Large N limit of superconformal field theories and supergravity}.
\newblock {\em Adv. Theor. Math. Phys.}, 2:231--252, 1998.

\bibitem{liu2011non}
Hong Liu, John McGreevy, and David Vegh.
\newblock {Non-Fermi liquids from holography}.
\newblock {\em Phys. Rev. D}, 83:065029, 2011.

\bibitem{faulkner2011emergent}
Thomas Faulkner, Hong Liu, John McGreevy, and David Vegh.
\newblock {Emergent quantum criticality, Fermi surfaces, and AdS(2)}.
\newblock {\em Phys. Rev. D}, 83:125002, 2011.

\bibitem{_ubrovi__2009}
Mihailo Cubrovic, Jan Zaanen, and Koenraad Schalm.
\newblock {String Theory, Quantum Phase Transitions and the Emergent Fermi-Liquid}.
\newblock {\em Science}, 325:439--444, 2009.

\bibitem{doi:10.1126/science.1189134}
Thomas Faulkner, Nabil Iqbal, Hong Liu, John McGreevy, and David Vegh.
\newblock {Strange metal transport realized by gauge/gravity duality}.
\newblock {\em Science}, 329:1043--1047, 2010.

\bibitem{Hartman_2010}
Thomas Hartman and Sean~A. Hartnoll.
\newblock {Cooper pairing near charged black holes}.
\newblock {\em JHEP}, 06:005, 2010.

\bibitem{Gubankova:2010ny}
E.~Gubankova.
\newblock {Particle-hole instability in the $AdS_4$ holography}.
\newblock In {\em {P- and CP-odd effects in hot and dense matter (CP-ODD)}}, 6 2010.

\bibitem{gubankova2011holographic}
E.~Gubankova, J.~Brill, M.~Cubrovic, K.~Schalm, P.~Schijven, and J.~Zaanen.
\newblock {Holographic fermions in external magnetic fields}.
\newblock {\em Phys. Rev. D}, 84:106003, 2011.

\bibitem{Ge:2011cw}
Xian-Hui Ge and Hong-Qiang Leng.
\newblock {Analytical calculation on critical magnetic field in holographic superconductors with backreaction}.
\newblock {\em Prog. Theor. Phys.}, 128:1211--1228, 2012.

\bibitem{Fang:2015dia}
Li-Qing Fang, Xiao-Mei Kuang, Bin Wang, and Jian-Pin Wu.
\newblock {Fermionic phase transition induced by the effective impurity in holography}.
\newblock {\em JHEP}, 11:134, 2015.

\bibitem{Jeong:2019zab}
Hyun-Sik Jeong, Keun-Young Kim, Yunseok Seo, Sang-Jin Sin, and Shang-Yu Wu.
\newblock {Holographic Spectral Functions with Momentum Relaxation}.
\newblock {\em Phys. Rev. D}, 102(2):026017, 2020.

\bibitem{Yuk:2022lof}
Taewon Yuk and Sang-Jin Sin.
\newblock {Flow equation and fermion gap in the holographic superconductors}.
\newblock {\em JHEP}, 02:121, 2023.

\bibitem{DeWolfe:2016rxk}
Oliver DeWolfe, Steven~S. Gubser, Oscar Henriksson, and Christopher Rosen.
\newblock {Gapped Fermions in Top-down Holographic Superconductors}.
\newblock {\em Phys. Rev. D}, 95(8):086005, 2017.

\bibitem{ghorai2024order}
Debabrata Ghorai, Taewon Yuk, and Sang-Jin Sin.
\newblock {Order parameter and spectral function in d-wave holographic superconductors}.
\newblock {\em Phys. Rev. D}, 109(6):066004, 2024.

\bibitem{ghorai2023fermi}
Debabrata Ghorai, Taewon Yuk, and Sang-Jin Sin.
\newblock {Fermi arc in p-wave holographic superconductors}.
\newblock {\em JHEP}, 10:003, 2023.

\bibitem{Fang:2014jka}
Li-Qing Fang, Xian-Hui Ge, Jian-Pin Wu, and Hong-Qiang Leng.
\newblock {Anisotropic Fermi surface from holography}.
\newblock {\em Phys. Rev. D}, 91(12):126009, 2015.

\bibitem{chakrabarti2022studyingholographicfermisurface}
Sayan Chakrabarti, Debaprasad Maity, and Wadbor Wahlang.
\newblock {Studying the holographic Fermi surface in the scalar induced anisotropic background}.
\newblock {\em Phys. Lett. B}, 827:136990, 2022.

\bibitem{Ling:2014bda}
Yi~Ling, Peng Liu, Chao Niu, Jian-Pin Wu, and Zhuo-Yu Xian.
\newblock {Holographic fermionic system with dipole coupling on Q-lattice}.
\newblock {\em JHEP}, 12:149, 2014.

\bibitem{Ling:2013aya}
Yi~Ling, Chao Niu, Jian-Pin Wu, Zhuo-Yu Xian, and Hong-bao Zhang.
\newblock {Holographic Fermionic Liquid with Lattices}.
\newblock {\em JHEP}, 07:045, 2013.

\bibitem{han2024mean}
Young-Kwon Han, Debabrata Ghorai, Taewon Yuk, and Sang-Jin Sin.
\newblock Mean field theory and holographic kondo lattice.
\newblock {\em arXiv preprint arXiv:2407.01978}, 2024.

\bibitem{yuk2024encoding}
Taewon Yuk and Sang-Jin Sin.
\newblock Encoding the lattice in the holography.
\newblock {\em arXiv preprint arXiv:2401.07498}, 2024.

\bibitem{li2012holographic}
Wei-Jia Li, Rene Meyer, and Hong-bao Zhang.
\newblock {Holographic non-relativistic fermionic fixed point by the charged dilatonic black hole}.
\newblock {\em JHEP}, 01:153, 2012.

\bibitem{wen2012dipole}
Wen-Yu Wen and Shang-Yu Wu.
\newblock {Dipole coupling effect of holographic fermion in charged dilatonic gravity}.
\newblock {\em Int.J.Mod.Phys.Conf.Ser}, 21:143--144, 2013.

\bibitem{song2019stability}
Geunho Song, Junchen Rong, and Sang-Jin Sin.
\newblock {Stability of topology in interacting Weyl semi-metal and topological dipole in holography}.
\newblock {\em JHEP}, 10:109, 2019.

\bibitem{Seo:2018hrc}
Yunseok Seo, Geunho Song, Yong-Hui Qi, and Sang-Jin Sin.
\newblock {Mott transition with Holographic Spectral function}.
\newblock {\em JHEP}, 08:077, 2018.

\bibitem{ghorai2024classes}
Debabrata Ghorai, Taewon Yuk, Young-Kwon Han, and Sang-Jin Sin.
\newblock {Classes of holographic Mott gaps}.
\newblock {\em JHEP}, 10:062, 2024.

\bibitem{Gursoy:2011gz}
Umut Gursoy, Erik Plauschinn, Henk Stoof, and Stefan Vandoren.
\newblock {Holography and ARPES Sum-Rules}.
\newblock {\em JHEP}, 05:018, 2012.

\bibitem{Faulkner:2010tq}
Thomas Faulkner and Joseph Polchinski.
\newblock {Semi-Holographic Fermi Liquids}.
\newblock {\em JHEP}, 06:012, 2011.

\bibitem{Smit:2021dwh}
S.~Smit et~al.
\newblock {Momentum-dependent scaling exponents of nodal self-energies measured in strange metal cuprates and modelled using semi-holography}.
\newblock {\em Nature Commun.}, 15(1):4581, 2024.

\bibitem{Mauri:2024wgc}
Enea Mauri, Steef Smit, Mark Golden, and H.~T.~C. Stoof.
\newblock {Gauge-gravity duality comes to the laboratory: Evidence of momentum-dependent scaling exponents in the nodal electron self-energy of cuprate strange metals}.
\newblock {\em Phys. Rev. B}, 109(15):155140, 2024.

\bibitem{Gubser:2009qt}
Steven~S. Gubser and Fabio~D. Rocha.
\newblock {Peculiar properties of a charged dilatonic black hole in $AdS_{5}$}.
\newblock {\em Phys. Rev. D}, 81:046001, 2010.

\bibitem{wu2011some}
Jian-Pin Wu.
\newblock {Some properties of the holographic fermions in an extremal charged dilatonic black hole}.
\newblock {\em Phys. Rev. D}, 84:064008, 2011.

\bibitem{gubser2012analytic}
Steven~S. Gubser and Jie Ren.
\newblock {Analytic fermionic Green's functions from holography}.
\newblock {\em Phys. Rev. D}, 86:046004, 2012.

\bibitem{ge2024thermo}
Xian-Hui Ge and Zhao-jie Xu.
\newblock {Thermo-electric transport of dyonic Gubser-Rocha black holes}.
\newblock {\em JHEP}, 03:069, 2024.

\bibitem{Ishigaki:2024pfv}
Shuta Ishigaki and Zhaojie Xu.
\newblock {Thermodynamics, magnetic properties, and global $U(1)$ symmetry breaking of the S-type Gubser-Rocha model}.
\newblock 6 2024.

\bibitem{ge2010analytical}
Xian-Hui Ge and Hong-Qiang Leng.
\newblock {Analytical calculation on critical magnetic field in holographic superconductors with backreaction}.
\newblock {\em Prog. Theor. Phys.}, 128:1211--1228, 2012.

\bibitem{iqbal2009real}
Nabil Iqbal and Hong Liu.
\newblock {Real-time response in AdS/CFT with application to spinors}.
\newblock {\em Fortsch. Phys.}, 57:367--384, 2009.

\bibitem{kittel2021introduction}
Charles Kittel.
\newblock Introduction to solid state physics eighth edition.
\newblock 2021.

\bibitem{PhysRevLett.72.709}
T.~Wang, K.~P. Clark, G.~F. Spencer, A.~M. Mack, and W.~P. Kirk.
\newblock Magnetic-field-induced metal-insulator transition in two dimensions.
\newblock {\em Phys. Rev. Lett.}, 72:709--712, 1994.

\bibitem{PhysRevLett.79.1543}
Dragana Popovi\ifmmode~\acute{c}\else \'{c}\fi{}, A.~B. Fowler, and S.~Washburn.
\newblock Metal-insulator transition in two dimensions: Effects of disorder and magnetic field.
\newblock {\em Phys. Rev. Lett.}, 79:1543--1546, 1997.

\bibitem{PhysRevB.104.045138}
Jingyao Meng, Rubem Mondaini, Tianxing Ma, and Hai-Qing Lin.
\newblock Inducing a metal-insulator transition in disordered interacting dirac fermion systems via an external magnetic field.
\newblock {\em Phys. Rev. B}, 104:045138, 2021.

\bibitem{PhysRevLett.57.1056}
V.~J. Goldman, M.~Shayegan, and H.~D. Drew.
\newblock Anomalous hall effect below the magnetic-field-induced metal-insulator transition in narrow-gap semiconductors.
\newblock {\em Phys. Rev. Lett.}, 57:1056--1059, Aug 1986.

\end{thebibliography}
\end{document}